\documentclass[fleqn,usenatbib]{mnras}

\usepackage{newtxtext,newtxmath}

\usepackage[T1]{fontenc}
\usepackage{inputenc}
\usepackage{ae,aecompl}

\usepackage{graphicx}	
\usepackage{amsmath}	
\usepackage{amssymb}	
\usepackage{tabularx} 

\usepackage{xcolor}

\graphicspath{{pictures/}} 

\title[Scalar fields and the S2 star]{Scalar field effects on the orbit of S2 star}

\author[The GRAVITY  Collaboration]{
The GRAVITY  Collaboration: 
A.	Amorim,$^{1,4}$
M.	Baub\"ock,$^5$
M.	Benisty,$^6$
J.-P. Berger,$^6$\newauthor
Y.	Cl\'enet,$^7$
V. Coud\'e du Forest,$^7$
T.	de Zeeuw,$^{8,5}$
J.	Dexter,$^5$
G.	Duvert,$^6$
A.	Eckart,$^{9,10}$\newauthor
F.	Eisenhauer,$^5$
Miguel C. Ferreira,$^1$\thanks{Corresponding author, e-mail: mcferreira@tecnico.ulisboa.pt}
F.	Gao,$^5$
Paulo J.V. Garcia,$^{1,2,3}$\thanks{Corresponding author, e-mail: pgarcia@fe.up.pt}
E.	Gendron,$^7$\newauthor
R.	Genzel,$^{5,11}$
S.	Gillessen,$^5$
P.	Gordo,$^{1,4}$
M.	Habibi,$^5$
M.	Horrobin,$^9$
A.	Jimenez-Rosales,$^5$\newauthor
L.	Jocou,$^6$
P.	Kervella,$^7$
S.	Lacour,$^{7,5}$
J.-B.	Le Bouquin,
P. L\'ena,$^7$
T.	Ott,$^5$
M.	P\"ossel,$^{12}$\newauthor
T.	Paumard,$^7$
K.	Perraut,$^6$
G.	Perrin,$^7$
O.	Pfuhl,$^5$
G.	Rodriguez Coira,$^7$
G.	Rousset,$^7$\newauthor
O.	Straub,$^5$
C.	Straubmeier,$^9$
E.	Sturm,$^5$
F.	Vincent,$^7$
S.	von Fellenberg,$^5$
I.	Waisberg$^5$\newauthor
and F.	Widmann$^5$
\\
$^1$CENTRA, Centro de Astrof\'{\i}sica e Gravita\c{c}\~{a}o, Instituto Superior T\'{e}cnico, Avenida Rovisco Pais 1, 1049 Lisboa, Portugal\\
$^{2}$Universidade do Porto, Faculdade de Engenharia, Rua Dr. Roberto Frias, 4200-465 Porto, Portugal\\
$^3$European Southern Observatory, Casilla 19001, Santiago 19, Chile\\
$^4$Universidade de Lisboa - Faculdade de Ci\^encias, Campo Grande, 1749-016 Lisboa, Portugal\\
$^5$Max Planck Institute for Extraterrestrial Physics (MPE), Giessenbachstr.1, 85748 Garching, Germany\\
$^6$Univ. Grenoble Alpes, CNRS, IPAG, 38000 Grenoble, France\\
$^7$LESIA, Observatoire de Paris, Universit\'e PSL, CNRS, Sorbonne Universit\'e, Universit\'e de Paris, 5 place Jules Janssen, 92195 Meudon, France\\
$^8$Sterrewacht Leiden, Leiden University, Postbus 9513, 2300 RA Leiden, The Netherlands\\
$^9$ Physikalisches Institut, Universit\"at zu K\"oln, Z\"ulpicher Str. 77, 50937 K\"oln, Germany\\
$^{10}$ Max-Plank-Institut f\"ur Radioastronomie, Auf dem H\"ugel 69, 53121 Bonn, Germany\\
$^{11}$Departments of Physics and Astronomy, Le Conte Hall, University of California, Berkeley, CA 94720, USA\\
$^{12}$Max Planck Institute for Astronomy (MPIA) and Haus der Astronomie, K\"onigstuhl 17, D-69117 Heidelberg, Germany}

\date{Accepted XXX. Received YYY; in original form ZZZ}

\pubyear{2015}

\begin{document}
\label{firstpage}
\pagerange{\pageref{firstpage}--\pageref{lastpage}}
\maketitle

\begin{abstract}
Precise measurements of the S-stars orbiting SgrA* have set strong constraints on the nature of the compact object at the centre of the Milky Way. The presence of a black hole in that region is well established, but its neighboring environment is still an open debate. In that respect, the existence of dark matter in that central region may be detectable due to its strong signatures on the orbits of stars: the main effect is a Newtonian precession which will affect the overall pericentre shift of S2, the latter being a target measurement of the GRAVITY instrument. The exact nature of this dark matter (e.g., stellar dark remnants or diffuse dark matter) is unknown.
This article assumes it to be an scalar field of toroidal distribution, associated with ultra-light dark matter particles, surrounding the Kerr black hole. 
Such a field is a form of "hair" expected in the context of superradiance, a mechanism that extracts rotational energy from the black hole. Orbital signatures for 
the S2 star are computed and shown to be detectable by GRAVITY. The scalar field can be constrained because the variation of orbital elements depends both on the relative mass of the scalar field to the black hole and on the field mass coupling parameter.
\end{abstract}




\newpage
\clearpage
\newpage

\section{Introduction}

\begin{table*}
\centering
\small
\caption{Literature computing "classical" GR and other effects on the orbits of the S-stars from Schwarzschild/Kerr black-holes. The following abbreviations are used: post-Newtonian (PN), orbital perturbation theory (PT), gravitational redshift (G-red), Newtonian precession (N-pre), pericentre precession (P-pre), Lense-Thirring precession (J-pre), quadrupole precession (Q-pre).}\label{tab:GR_literature}
\begin{tabularx}{\textwidth}{llXXX} 
\hline
metric & GR effects & other effects & observables & reference \\\hline
Kerr &  P-pre, J-pre & lensing     & time averaged orbital effects & \cite{Jaroszynski1998} \\ 
Schwarzschild & G-red, P-pre, J-pre  & none     & astrometric orbital fitting, time averaged orbital effects, spectroscopic  & \cite{Fragile2000} \\
PN & P-pre & N-pre from extended mass & astrometric orbits  &\cite{Rubilar2001} \\
PN & P-pre, J-pre & N-pre from extended mass, stellar remnants & time averaged orbital effects  &\cite{Weinberg2005} \\
PN & G-red  & no     & astrometric orbits and spectroscopic  &\cite{Zucker2006} \\
Kerr & P-pre, J-pre &Kerr-de Sitter gravitational field     & time averaged orbital effects &\cite{Kraniotis2007} \\ 
PN & P-pre, J-pre, N-pre &stellar cluster     & pericentre shift &\cite{Nucita2007} \\ 
PN & P-pre, J-pre, Q-pre & testing "no-hair" theorem & time averaged orbital effects & \cite{will2008testing} \\ 
Kerr & P-pre, J-pre &no   & integrated orbital effects, spectroscopic & \cite{Kannan2009} \\ 
PN & P-pre, J-pre &N-pre from extended mass   &general astrometric, spectroscopic & \cite{Preto2009} \\ 
PN & G-red, P-pre, J-pre  & light-path effects  &spectroscopic & \cite{Angelil2010a} \\ 
PN & G-red, J-pre, Q-pre  & pulsar timing  &spectroscopic & \cite{Angelil2010b} \\ 
PN & P-pre, J-pre, Q-pre & N-pre from extended mass  & astrometric & \cite{Merritt2010} \\ 
PN & P-pre, J-pre, Q-pre & N-pre from extended mass  & spectroscopic & \cite{Iorio2011a} \\ 
PN & P-pre, J-pre, Q-pre & gravitational waves  & time averaged orbital effects & \cite{Iorio2011b} \\ 
PT& J-pre, Q-pre & perturbing effects of cluster stars  & time averaged orbital effects & \cite{Sadeghian2011} \\ 
Kerr  &full GR & full lensing    &astrometric, spectroscopic, black hole spin measurement  & \cite{Zhang2015}  \\    
PN & P-pre, J-pre, Q-pre  & N-pre, EHT, pulsars, "no-hair" theorem    & astrometric  & \cite{Psaltis2016}  \\
Kerr & full GR  & full lensing    & astrometric, spectroscopic, black hole spin inclination effects  & \cite{Yu2016}  \\     
Kerr   & full GR & lensing primary & astrometric, spectroscopic & \cite{grould2017general}  \\       
PN & P-pre, J-pre, Q-pre  & effect of orbital eccentricity  & periods evolution &\cite{Iorio2017a}  \\
PN & P-pre, J-pre  &  &spectroscopic, time averaged orbital effects  &\cite{Iorio2017b}  \\
PN & P-Pre &&astrometric&\cite{Parsa2017}\\
Kerr & full GR & full lensing, Newtonian perturbations  & astrometric, spectroscopic, time averaged orbital effects &\cite{Zhang2017a}\\
\hline
\end{tabularx} 
\end{table*}

SgrA* is the nearest putative supermassive black hole and a unique laboratory to study gravity, compact objects and dark matter \citep[e.g.,][]{Genzel2010,Johannsen2016,Alexander2017,Do2019}. It is therefore a prime target for many current facilities such as the Event Horizon Telescope \citep[EHT, e.g.,][]{Broderick2014,Broderick2016, Lu2018,Pu2018}, the Global Millimeter VLBI Array \citep{Issaoun2019}, the VLTI/GRAVITY \citep{Gravity2017, Gravity2018a, Gravity2018b,Gravity2019} and future facilities such as the ELTs \citep{Weinberg2005, Trippe2010}, the SKA \citep{Bull2018} or LISA \citep{Gourgoulhon2019}. Methods probing the central compact object include radiation signatures \citep[e.g.,][]{Eckart2006,Doeleman2008,Broderick2011}, pulsar timing \citep[e.g.,][]{Liu2012, Zhang2017b},  gravitational waves or the motion of test particles \citep[stars, e.g.,][]{Gillessen2017, Jia2019}. 

The study of the motion of test particles around SgrA* focused on the inner stars (the S-stars),  using both spectroscopy and astrometry.  Before the \cite{Gravity2018a} breakthrough all astrometric and spectroscopic measurements were  fitted by Newtonian physics (i.e., a Keplerian orbit), allowing the extraction of the black hole mass (as well as the distance). However, through the years, many predictions were developed to test gravity in this environment. They either focused on  General Relativity (GR)  or on extensions/alternatives to GR.

The predictions of GR effects for the orbits of the S-stars, and in particular S2, addressed several aspects: gravitational redshift, pericentre shift,  Lense-Thirring effect, quadrupole moment, "hair" in the black hole, lensing effects on the photons, Shapiro time delay. Furthermore, Newtonian effects from nearby stars and extended (dark) mass distributions, which could contaminate the GR signatures were studied.
The effects described above impact on the astrometric position and/or the spectroscopic line of sight velocity of the S-stars. In Table~\ref{tab:GR_literature} the literature addressing these GR (\& other) effects in the orbits of the S-stars is summarised.\footnote{ Astrophysical effects of hydrodynamical origin on the S-stars, from the local plasma, or their stellar winds, are negligible \citep{Psaltis2012,Psaltis2013}.} \cite{Gravity2018a} detected the first GR effect in the orbit of the S2 star -- gravitational redshift -- at 10 $\sigma$, this result was later refined at 20 $\sigma$ by \cite{Gravity2019} and recently confirmed at 5 $\sigma$ by \cite{Do2019b} using different instrumentation. Observational work is ongoing towards the detection of the pericentre shift of S2 and the discovery of putative closer stars, which could allow an astrometric measurement of the black hole spin \citep[e.g.,][]{Waisberg2018}.

\begin{table*}
\centering
\small
\caption{Literature computing extensions/alternatives to GR effects in the orbits of the S-stars.}\label{tab:eGR_literature}
\begin{tabularx}{\textwidth}{XXX} 
\hline
extension/alternative & results/comments &  reference \\\hline
charged non-rotating black holes&  Upper limit to black hole charge from S2 precession upper limit.& \cite{DeLaurentis2018a}, \cite{Iorio2012}, \cite{Zakharov2018b} \\
charged rotating black holes and plasma effects& upper limits from black hole mass, spin and local magnetic field&\cite{Zajacek2018}\\
fermion ball& Ruled out by \cite{Ghez2005} and \cite{Gravity2018a}.&\cite{Munyaneza2002}\\
boson "star"&Effects much smaller than GR at S2 orbit, only relevant at a few tens of Schwarzschild radii.&\cite{Amaro-Seoane2010},\cite{Boshkayev2019},\cite{Grould2017} \\
Yukawa potential&Upper limits on potential parameters and graviton mass from S2 precession upper limit.&\cite{Borka2013}, \cite{Hees2017}, \cite{Zakharov2016}, \cite{Zakharov2018a} \\
Einstein-Maxwell-Dilaton-Axion gravity&Effects smaller than $10^{-3}$ of GR for S2, need pulsars or inner stars for further tests.&\cite{DeLaurentis2018a}\\
Brans-Dicke theory&Effects smaller than $10^{-3}$ of GR for S2, need pulsars or inner stars for further tests.&\cite{DeLaurentis2018a}, \cite{Kalita2018}\\
$f(R)$ gravity&Effects smaller than $10^{-3}$ of GR for S2, need pulsars or inner stars for further tests.&\cite{Capozziello2014}, \cite{DeLaurentis2018a}, \cite{DeLaurentis2018b}, \cite{Kalita2018}\\
nonlocal gravity& Precession compatible with observational upper limit, of the order of GR prediction&\cite{Dialektopoulos2019}\\
scalar tensor gravity& Precession is $13\times$ GR value, ruled out by \cite{Hees2017}.&\cite{Borka2019}\\
$f(R,\phi)$ gravity&Best fit precession prediction for S2 is $20\times$ GR value, ruled out by  \cite{Hees2017}.&\cite{Capozziello2014}\\
hybrid gravity&Best fit precession prediction too high, ruled out by  \cite{Hees2017}.&\cite{Borka2016}\\
$R^n$ gravity&When compared with \cite{Hees2017} upper value, the GR value ($n=1$) is recovered to  $<1\%$, or smaller if extended mass distributions are present.&\cite{Borka2012}, \cite{Zakharov2014}\\
quadratic Einstein-Gauss-Bonnet gravity & Derive expressions for gravitational redshift in function of theory coupling parameters (scalar/matter \& scalar/Gauss-Bonnet invariant).&\cite{Hees2019}\\
dark matter profiles (See Table~\ref{tab:GR_literature} for dark matter + black hole studies.)&Dark matter mass required to explain TeV emission compatible with orbital upper limits. Limits on spatial distribution of non-annihilating dark matter.&\cite{dePaolis2011}, \cite{Dokuchaev2015}, \cite{Hall2006}, \cite{Iorio2013}, \cite{Lacroix2018},\cite{Zakharov2007}\\
scalar fields and ultralight dark matter &Upper limits on scalar field mass (1\% of black hole) for particles of  mass $4\times 10^{-19}~\mathrm{eV/c}^2$&\cite{Bar2019}\\
\hline
\end{tabularx} 
\end{table*}

Extensions/alternatives to GR in the context of S-stars orbits were also developed\footnote{Extensions/alternatives making other types of predictions such as lensing and/or electromagnetic radiation (shadows, annihilation) are not addressed in this paper.}. There are several arguments (related to dark energy, dark matter and unification)  why GR shouldn't be the final word on gravity \citep[e.g., ][]{Berti2015,Cardoso2019} and on caution on interpreting observations as proofs of GR black holes \citep[e.g.,][]{Abramowicz2002, Cardoso2017,Mizuno2018}. Still, the case for a supermassive black hole at the Galactic Centre is extremely strong, and this hypothesis has passed much more tests than its alternatives \citep{Eckart2017}. Alternatives with astrometric signatures in the literature can be grouped as: a) "classical" GR charged black-holes; b) dark matter profiles (including fermion balls \& boson "stars"); c) several variations of GR. These are summarized in Table~\ref{tab:eGR_literature}. It emerges that orbital precession is a strong falsifier of theories and a critical test of their validity. Some theories can't be tested by the S2 orbit because they asymptotically match GR at scales much smaller than S2. Extended dark matter distributions surrounding the black hole are among those with stronger signatures and more stringent limits.

In this paper we address a hybrid scenario of a Kerr black hole with "hair" in the form of a scalar field. Scalar fields appear at the meeting point between phenomenological necessity and theoretical consistency. Given that they are very simple objects to manipulate, scalar fields are introduced \textit{ad hoc} in several domains of physics. One of the most famous cases is the axion, the scalar field that was introduced by \cite{peccei1977cp} to solve the strong CP problem. The theoretical investigations of high energy theories, of which String Theory is an example, require that the low energy effective models contain a set of scalar fields with a very small, but non-vanishing rest mass \citep{svrcek2006axions}. Given the theoretical resemblance between this set of fields and the original Peccei-Quinn's axion, all of these proposed scalar fields are generically called axions, and the scenario of their putative existence is called the ``Axiverse'' \citep{arvanitaki2010string}, in which the masses of the scalar fields can be as small as $m_s \sim 10^{-33}\text{ eV}$. Such ultra-light scalar fields are expected, depending on their mass scale, to leave phenomenological imprints both at cosmological scale \citep[see][and references therein]{arvanitaki2010string,marsh2016axion,hui2017ultralight} 
and astrophysical scale\footnote{For instance, \cite{isern2018axions} studied the effect of axions on the luminosity of white dwarfs, \cite{widdicombe2018formation} explore the formation of axion stars and \cite{baumann2019probing} study the hypothesis of scalar field clouds affecting the dynamics of binary black holes.}. Our main interest will be astrophysical, particularly on the interaction between scalar fields and black holes.

In the vicinity of astrophysical black holes, massive scalar fields develop quasi-bound state solutions\footnote{\cite{herdeiro2014kerr} constructed exact bound state solutions of a scalar field in equilibrium with a BH.} \citep[see e.g.,][]{detweiler1980klein, dolan2007instability, witek2013superradiant}, i.e., configurations of the scalar field that vanish at infinity and correspond to ingoing waves at the black hole event horizon. Some of these solutions decay with time, being radiated away to infinity, but if the Compton wavelength of the scalar field is comparable with the gravitational radius of the black hole, it is possible to find growing modes. In these situations, the quasi-bound states do not decay and can have a long-lived existence maintained by a slow, but constant, extraction of rotational energy from the black hole. The phenomenon behind this extraction of energy is called superradiance (see \cite{brito2015superradiance} and references therein). In the case of the black hole in the Galactic Centre, with mass given by $M_{\bullet} \sim 10^{6} M_{\odot}$, the value of the rest mass of the scalar field for which the bound states can engage in a superradiant energy extraction is given by \cite{kodama2012axiverse} who find $m_s \lesssim 10^{-17}\text{ eV}$ . The existence of these scalar-field bound states highlights the possibility that BHs can capture and maintain in its vicinity (for astrophysically-relevant periods of time) scalar-field structures that result from scattering events \citep[see e.g.,][]{dolan2013superradiant,witek2013superradiant,okawa2014black}. This motivates the study of the astrophysical effects of the existence of scalar field bound states \citep[e.g.,][]{brito2015black, cunha2015shadows, vincent2016astrophysical, rosa2018stimulated}, particularly their possible influence on the orbits of stars around black holes \citep{ferreira2017orbital, fujita2017ultralight}.

Our approach is to consider a scenario in which a scalar field bound-state structure has developed around the black hole in the centre of our galaxy and investigate how such a structure influences the orbit of the star S2. We will make a theoretical analysis of this scenario and leave for future work a full fit of available data. In Section~\ref{subsec:peak-of-cloud} the scalar field is introduced. Because of its fractional mass to the black hole ($\Lambda$) is very small it can be described by a potential perturbing the orbit of S2 (cf. Sections~\ref{sec:describing} and Sections~\ref{sec:perturbing}). In Section~\ref{sec:results} the integrated effects over one orbit are computed, as function of the scalar field structure's fractional mass and mass coupling parameter  ($\alpha$). It is found that, as expected, the strongest effects on the S2 orbit take place when its orbit crosses the scalar field peak density regions. Depending on the $\alpha$ parameter and black hole inclination the scalar field can produce prograde (GR-like) or retrograde (Newtonian-like) pericentre precession. The intensity of the scalar field effects scale linearly with the relative mass of the structure and non-linearly with the $\alpha$ parameter, crossing the GR expected values. Besides precession the field also produces inclination and ascending node variations, with amplitudes that may compete with the Lense-Thirring effect. Other parameters, such as eccentricity, are also found to vary. These results are discussed in Section~\ref{sec:discussion}. Several details of the calculations are presented in appendices.

\section{Methods}\label{sec:methods}

\subsection{The scalar field structure}\label{subsec:peak-of-cloud}

The starting point of our study is the possibility that scalar fields develop astrophysically relevant structures around black holes. To study this possibility, we will analyse the solutions to the Klein-Gordon equation in a Kerr space time. We will follow the analytic results of \cite{detweiler1980klein} and then translate the scalar field solution in an effective gravitational potential which can then be treated with the usual perturbation analysis of Keplerian orbits. In this section we will be using Planck units ($\hbar = c = G = 1$) unless otherwise stated.

A black hole-scalar field system, in which the scalar field is minimally coupled to gravity, is described by the following action
\begin{equation}
  S = \int d^4x \sqrt{-g} \left(\frac{R}{16\pi} - \frac{1}{2} g^{\alpha \beta} \Psi^*_{,\alpha} \Psi^*_{,\beta} - \frac{\mu^2}{2} \Psi \Psi^*\right)
\end{equation}
in which $R$ is the Ricci scalar, $g_{\mu\nu}$ and $g$ is the metric and its determinant, $\Psi(t,r,\theta,\phi)$ is a complex scalar field\footnote{We choose to deal with the complex scalar field but the real scalar field can also be considered, as in \cite{brito2015black} and \cite{ferreira2017orbital}} and $\mu$ is the mass of the scalar field. The principle of least action results in the Einstein-Klein-Gordon system of equations
\begin{equation}
\begin{cases}
  G^{\alpha\beta} =& 8\pi T^{\alpha\beta}\\
  \nabla_{\alpha}\nabla^{\alpha}\Psi =& \mu^2\Psi
\end{cases}
\end{equation}
where $G_{\alpha\beta}$ is the Einstein tensor, $\nabla_{\alpha}$ represents the covariant derivative and
\begin{equation}
  T^{\alpha\beta} = \Psi^{*,(\alpha}\Psi^{,\beta)} - \frac{1}{2} g^{\alpha\beta} \left(\Psi^*_{,\sigma}\Psi^{,\sigma} + \mu^2 \Psi^* \Psi\right)
\end{equation}
is the energy-momentum of the scalar field. In this system, the relevant quantity is the dimensionless mass coupling given by
\begin{equation}
  \alpha = r_g \mu = \left[\frac{GM}{c^2}\right] \left[\frac{m_s c}{\hbar}\right] = \frac{M_{\bullet} m_s}{m_P^2},
\end{equation}
using fundamental constants for physical clarity, where $r_g$ is the gravitational radius, $\lambda_C = \mu^{-1}$ is the Compton wavelength of the particle with mass $m_s$ and $m_P$ is the Planck mass. In Plank units it is usually written as
\begin{equation}
  \alpha = \mu M_{\bullet}.
\end{equation}
Considering that the influence of the black hole dominates the spacetime, the Klein-Gordon equation can be solved assuming a fixed Kerr metric as the background. In this case, one can take the limit $\alpha \ll 1$ and obtain that the time-dependence of the quasi-bound-state solutions is (see Appendix~\ref{app:BH-SF-setup})
\begin{equation}
    \Psi \sim \mathrm{e}^{-\mathrm{i} \omega t}, \quad \omega = \omega_R + \mathrm{i} \omega_I
\end{equation}
where the frequency is a sum of a real and an imaginary component, both positive. The existence of a positive imaginary component of the frequency means that there will be an exponential growth with time of the scalar field profile. This growth is characterized by the time scale $\tau_I = 1/\omega_I$ which is given by $\tau_I \sim \alpha^{-9}$ for the fastest growing mode\footnote{A value of $\alpha \sim 10^{-2}$ in the context of the BH in the center of the galaxy corresponds to $\tau_I \sim 10^3 \text{ Gyr}$.}. However, the oscillation of the scalar field profile, expressed by the presence of the real part of the frequency $\omega_R$, has a time scale $\tau_R = 1/\omega_R$ which is much smaller than the growth time scale, more explicitly $\tau_R \sim \alpha^{-1} \ll \tau_I$. This observation means that one can separate the timescales and consider that the dynamics of the growth of the scalar profile can be ignored if one focus on the dynamics of its oscillation. This assumption, which has been used in, e.g., \cite{brito2015black} and \cite{ferreira2017orbital}, amounts to consider that the scalar field profile is given by
\begin{equation}\label{eq:Psi_main}
  \Psi =  A_{0} \mathrm{e}^{-\mathrm{i} (\bar{\omega}_R \bar{t} - \phi)}  \bar{r} \alpha^2 \mathrm{e}^{-\frac{\bar{r} \alpha^2}{2}} \sin\theta,
\end{equation}
for a given constant $A_0$. This constant is related to the total mass of the scalar cloud, which is considered to be constant in time scales we are focusing on. In the above equation coordinates normalized by the black hole mass are used
\begin{equation}
  \label{eq:normalized-coordinates}
  \begin{cases}
    r &\rightarrow \bar{r} = r/M_{\bullet}\\
    t &\rightarrow \bar{t} = t/M_{\bullet}
  \end{cases}.
\end{equation}

By considering that the field is mainly described by the fastest growing mode, the total mass of the scalar cloud is given by
\begin{equation}
\label{eq:mass-cloud-eq}
  M_{\mathrm{cloud}} = \int \rho r^2 \sin \theta dr d\theta d\phi
\end{equation}
where $\rho = T_{00}$ which in the limit $\alpha \ll 1$ is well described by (see Appendix~\ref{app:einstein-to-poisson})
\begin{equation}
  \label{eq:scalar-density}
  \rho \sim \mu^2 \left|\Psi\right|^2 =\mu^2 \left( A_0^2 \alpha^4 \mathrm{e}^{- \alpha^2 \bar{r}} \bar{r}^2 \sin^2\theta\right).
\end{equation}
We can solve the integral for the total mass of the cloud and obtain
\begin{equation}
  M_{\mathrm{cloud}} = \frac{64 \pi A_0^2}{\alpha^4} M_{\bullet}.
\end{equation}
From here we can see that the growth of the scalar field profile corresponds to an increase in the mass of the scalar cloud. 

The scalar field density distribution described by Eq.~\eqref{eq:scalar-density} can be characterized by an effective peak position  $R_\mathrm{peak}$ and effective width $\Delta C$. These can be estimated as
\begin{align}\label{eq:peak-value}
     R_\mathrm{peak} &= \langle \bar{r}\rangle =  \dfrac{\int_0^\infty \rho \bar{r} d\bar{r}}{\int_0^\infty \rho d\bar{r}} = \dfrac{3}{\alpha^2},\\
     \Delta C &= 2 \sqrt{\langle \bar{r}^2\rangle-\langle \bar{r}\rangle^2 } =\dfrac{2 R_\mathrm{peak}}{\sqrt{3}}\sim R_\mathrm{peak}.
\end{align}
One sees that the dimensionless mass coupling $\alpha$ dictates the position and width of the scalar field cloud. For large mass couplings the cloud is located close to the black hole and has a small effective width. For small mass couplings the cloud is located further away and has a larger width. The black hole spin value has no effect on the cloud location, however the black hole spin orientation determines the cloud central axis.

\begin{figure*}
  \begin{center}
  \includegraphics[width=1\columnwidth]{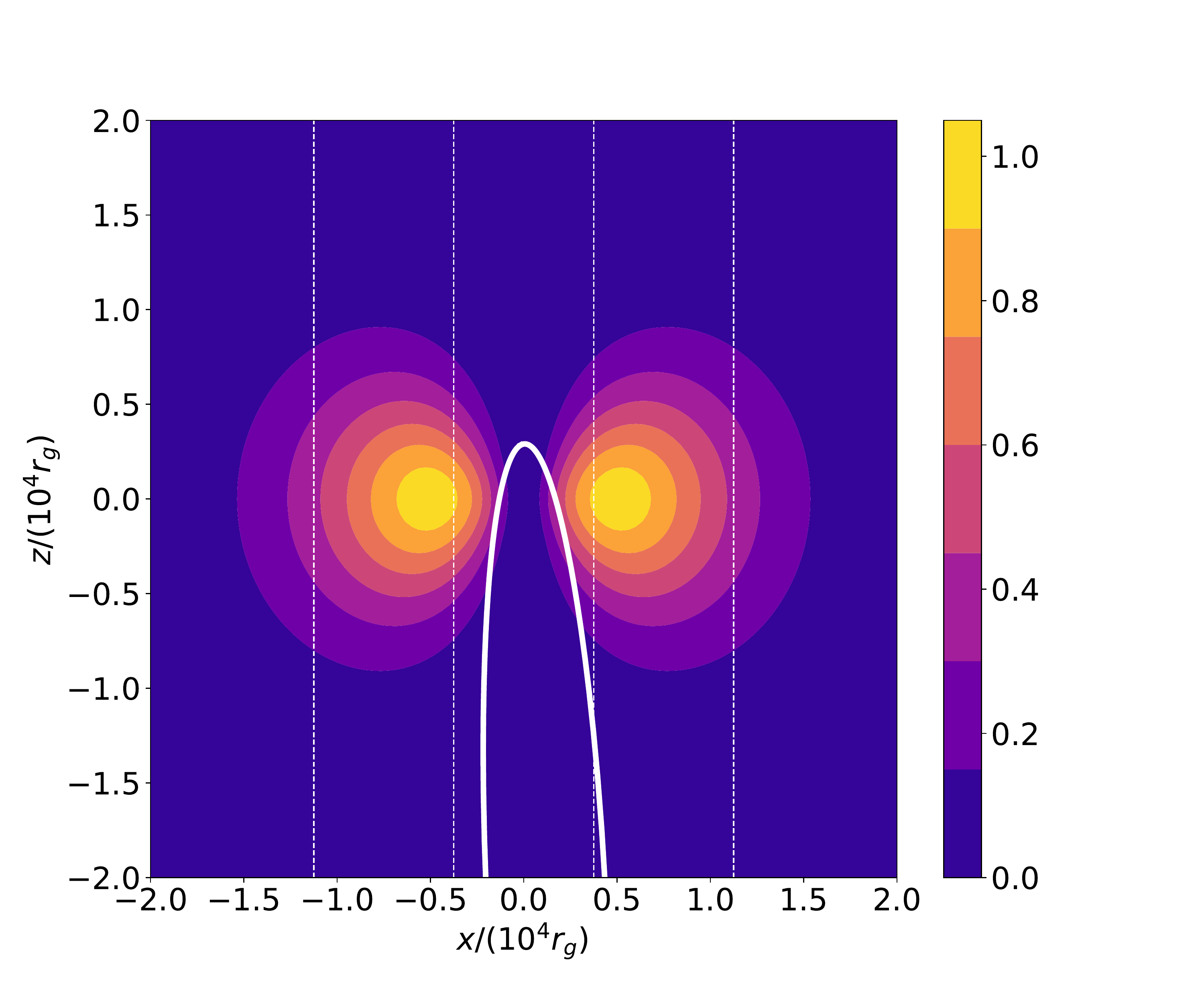}
  \includegraphics[width=1\columnwidth]{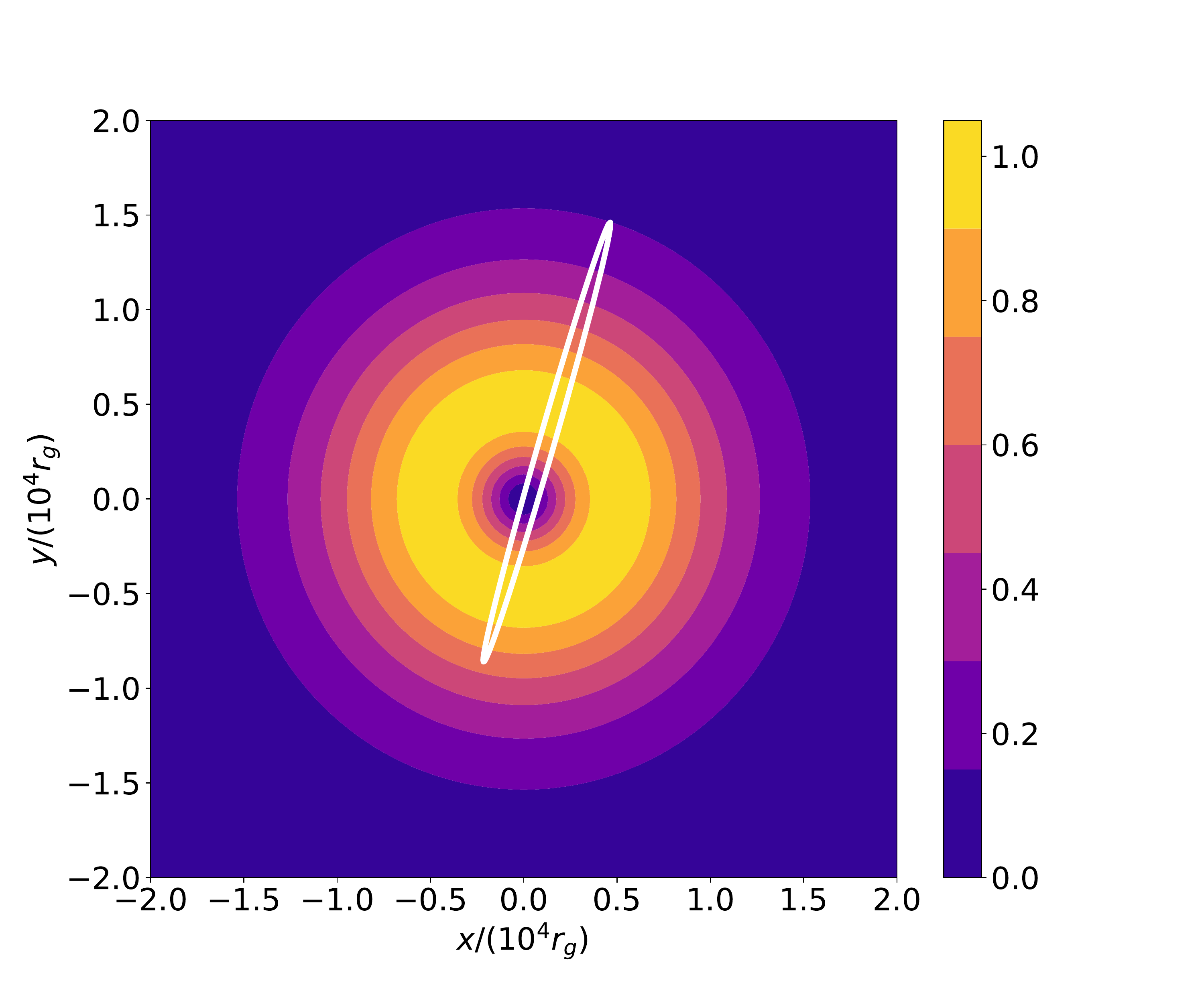}
  \end{center}
  \caption{Normalized density distribution of the scalar field cloud,  in the black hole reference frame.  If the scalar field cloud is to have a significant influence on the orbit of the star, the latter must intercept the former. An orbit with parameters  given by \citet[][cf. also equation~\eqref{eq:orbital-parameters}]{Grould2017}  is depicted by thick white curve. Left: $xz$ cut across the density distribution. The dashed lines indicate the regions  $\pm(R_\mathrm{peak}\pm \Delta C)$ defined by equations~\eqref{eq:peak-value}.
  Right: $xy$ cut across the density distribution, including the orbit projection.}
  \label{fig:scheme_cartoon}
\end{figure*}

\subsection{Describing the effect of the scalar field}\label{sec:describing}

Considering that the scalar field cloud can be described in terms of a Newtonian gravitational potential, we will calculate it and describe how it affects the orbits around the black hole.

The analytical profile of the scalar field we are using (cf. Eq.~\eqref{eq:scalar-density}) is valid in the limit $\alpha \ll 1$, which implies that
\begin{equation}
    r_g \ll R_{\mathrm{peak}},
\end{equation}
meaning the the scalar field structure attains its maximum far from the gravitational radius of the black hole. This allows us to consider that, in the region where the scalar cloud peaks  one can consider it as a perturbation to a flat background. One can, then, conceive a scenario in which the scalar field cloud that develops around the black hole in the Galactic Centre corresponds to a sort of toroidal density distribution that is schematically represented in Fig.~\ref{fig:scheme_cartoon}. 

To describe the gravitational potential that results from the presence of the scalar field cloud in a region far from the black hole we solve  Poisson's equation (see Appendix~\ref{app:einstein-to-poisson})
\begin{equation}
  \bar{\nabla}^2 U_{\mathrm{sca}} = - 4\pi(M_{\bullet} \mu)^2 |\Psi|^2,
\end{equation}
where the bar over the $\nabla$-operator means that differentiation is taken with respect to the normalized coordinates of Eq.~\eqref{eq:normalized-coordinates}. It can be rewritten, more explicitly as
\begin{equation}
\label{eq:poisson-equation-barred-variables}
  \bar{\nabla}^2 U_{\mathrm{sca}} = - 4\pi\left[\frac{M_{\mathrm{cloud}}}{M_{\bullet}}\right]\left(\frac{\alpha^{10}}{64\pi} \mathrm{e}^{- \alpha^2 \bar{r}} \bar{r}^2 \sin^2\theta\right).
\end{equation}
To solve this equations we use the harmonic decomposition technique to obtain an expression for the gravitational potential that can be written as (see Appendix~\ref{app:solving-the-potential})
\begin{equation}
  \label{eq:grav-potential-planck}
  U_{\mathrm{sca}} = \Lambda \left[ P_1(\bar{r}) + P_2(\bar{r})\cos^2\theta\right]
\end{equation}
with $\Lambda = M_{\mathrm{cloud}}/M_{\bullet}$ being the fractional mass of the scalar field cloud to to the black hole mass and
\begin{align}
  P_1(\bar{r}) &= \frac{16 \alpha^4 \bar{r}^2+48}{16 \alpha^4 \bar{r}^3}-\frac{e^{-\alpha^2 \bar{r}}}{16 \alpha^4 \bar{r}^3} \Big[\alpha^{10} \bar{r}^5\nonumber\\
         &+6 \alpha^8 \bar{r}^4+20 \alpha^6 \bar{r}^3+40 \alpha^4 \bar{r}^2+48 \alpha^2 \bar{r}+48\Big]\,,\\
  P_2(\bar{r}) &= -\frac{9}{\alpha^4 \bar{r}^3} + \frac{e^{-\alpha^2 \bar{r}}}{16 \alpha^4 \bar{r}^3} \Big[\alpha^{10} \bar{r}^5+6 \alpha^8 \bar{r}^4\nonumber\\
         &+24 \alpha^6 \bar{r}^3+72 \alpha^4 \bar{r}^2+144 \alpha^2 \bar{r}+144\Big].
\end{align}

\subsection{Perturbing the orbit of S2}\label{sec:perturbing}

For simplicity we will stop using barred quantities. The equations of motion governing the behavior of a star around the black hole surrounded by a scalar cloud is given by
\begin{equation}\label{eq:Newton_perturbed}
  \frac{d^2 \vec{r}}{dt^2} = -\frac{\vec{r}}{r^3} + \vec{F}_{\mathrm{pert}}
\end{equation}
where $\vec{r}$ is the normalized point-mass position vector with respect to the black hole.
In solving this problem, we consider that the gravitational potential due to the scalar cloud acts as a perturbation of a Keplerian orbit. To compute its  effect on the S2 star, we will have to use the Gauss equations (see Appendix~\ref{app:keplerian-orbits-formalism}) for the parameters characterizing its orbit.

The perturbing force that results from the presence of the scalar cloud is given by
\begin{equation}
  \vec{F}_{\mathrm{pert}} = \Lambda \nabla \left[ P_1(r) + P_2(r)\cos^2\theta\right]
\end{equation}
and can be decomposed as (see Appendix~\ref{app:keplerian-orbits-formalism})
\begin{align}
  F_R/\Lambda &= \sin ^2(i) \sin ^2(f+\omega ) P_2'(r)+P_1'(r),\\
  F_T/\Lambda &=-\frac{\sin ^2(i) (e \cos (f)+1) \sin (2 (f+\omega )) P_2(r)}{a \left(e^2-1\right)},\\
  F_N/\Lambda &=-\frac{\sin (2 i) (e \cos (f)+1) \sin (f+\omega ) P_2(r)}{a \left(e^2-1\right)},
\end{align}
where the prime $ '$ stands for derivative with respect to the radial coordinate, $f$ is the true anomaly and $a,e,i,\omega$ are the semi-major axis, the eccentricity, the inclination and the argument of the pericentre, respectively.

\subsection{The orbital elements of the orbit}

The framework we set up up until now is developed in a reference frame which is centered on the black hole and whose $z$-axis is aligned with the black hole's spin direction. The visual orbital parameters of the S2-star must be projected in such a reference frame. One can obtain them from the measured, Earth-based reference frame values in \cite{Gravity2018a,Gravity2019} by applying a set of rotations that relate the two frames (see \cite{grould2017general} for a detailed description). However, given the uncertainty in the orientation of the black hole's spin, the aforementioned conversion is not well defined. Facing this problem, we decided to, in a first run of our calculations, use the orientation proposed by \cite{grould2017general}. 

The orbital elements for the orbit of the S2 star in the re-scaled units read (\cite{Gravity2018a,Gravity2019})
\begin{align}
  &a_0 = 2.5 \times 10^4, \quad e_0 = 0.88473, \quad i_0 = 133.817^{\circ} \nonumber\\
  &\omega_0 = 66.12^{\circ}, \quad \Omega_0 = 227.82^{\circ},
\end{align}
which correspond, in the black hole-centered reference frame defined in \cite{grould2017general}, to
\begin{align}
\label{eq:orbital-parameters}
  &a_0 = 2.5 \times 10^4, \quad e_0 = 0.88473, \quad i_0 = 90.98^{\circ} \nonumber\\
  &\omega_0 = 81.60^{\circ}, \quad \Omega_0 = 254.191^{\circ}.
\end{align}

\subsection{Calculating the orbital elements variations}

One can calculate the average variation of the orbital elements of S2 over one period using the standard integral
\begin{equation}
\label{eq:Delta-kappa}
  \langle\Delta \kappa\rangle = \int_{f_0}^{f_0 + 2\pi} \frac{d \kappa}{dt}\frac{dt}{df'}df'.
\end{equation}
where $\kappa \in \{a,e,i,\Omega,\omega, \mathcal{M}_0\}$ are the usual elements (cf. Appendix~\ref{app:keplerian-orbits-formalism} Eqs.~\eqref{eq:kappa-1} to \eqref{eq:kappa-6}).  $dt/df'$ is obtained by inverting an embodiment of the Kepler equation
\begin{equation}
  \frac{df}{dt} = \sqrt{\frac{1}{a^3}}\left(\frac{(1 + e \cos f)^2}{(1 - e^2)^{3/2}}\right).
\end{equation}

\section{Results} \label{sec:results}

In this section the mean variation of the orbital parameters over a complete orbit are presented. These variations will be related to the mass coupling parameter $\alpha$ and relative scalar field mass $\Lambda$. As it clear from Section~\ref{subsec:peak-of-cloud}, the distribution of the scalar field density varies quite dramatically with the value of $\alpha$. This is crucial because one expects that the effects on the orbit of the star will depend on the position of the scalar field cloud with respect to it.

\begin{figure*}
\includegraphics[width=2\columnwidth]{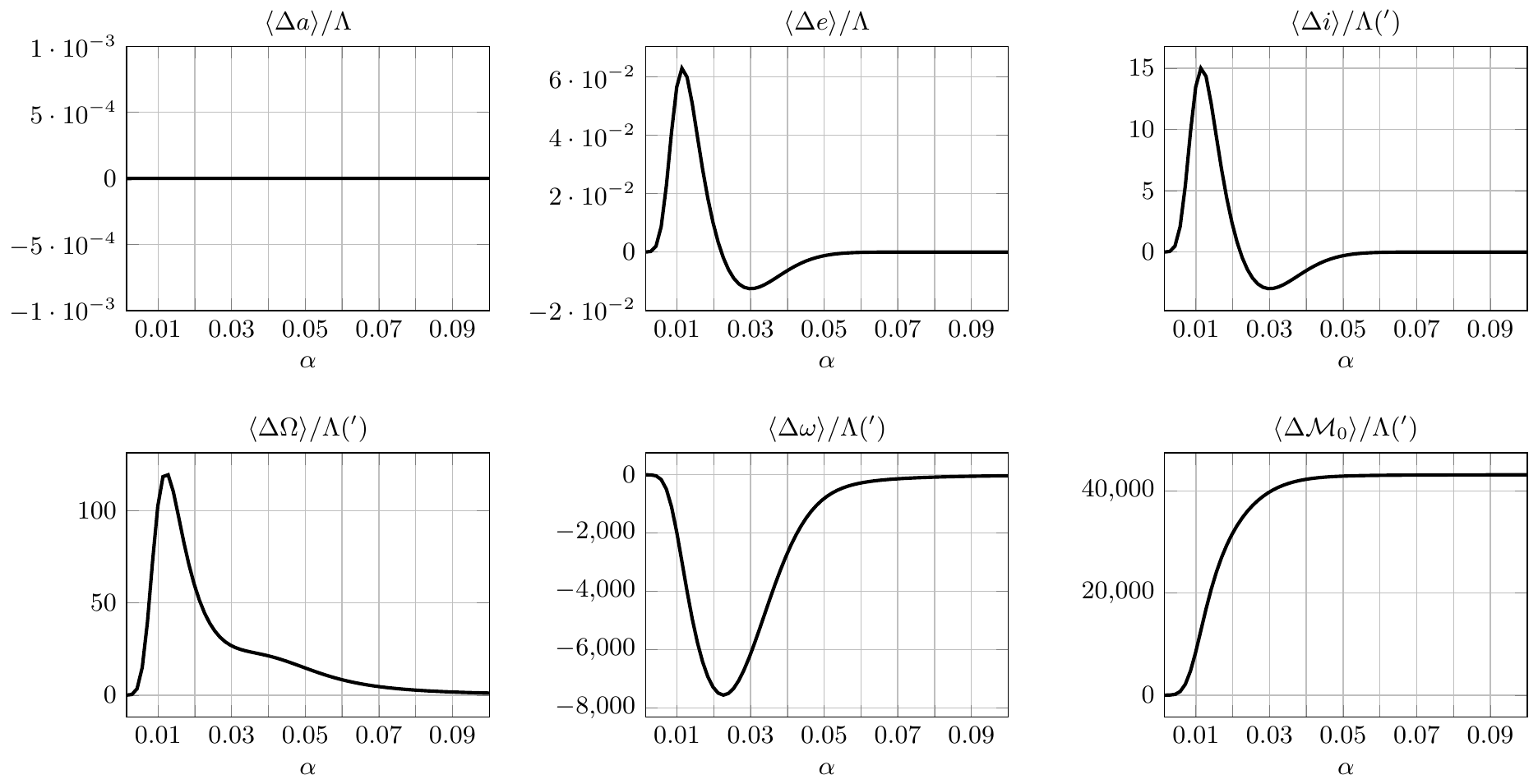}\\
\caption{Average variation of the orbital parameters over a period of the star S2 resulting from the presence of a scalar field cloud. The black hole-centered reference frame considered in doing these calculations is defined by \protect\cite{grould2017general}. Notice that the angular elements are presented in arcminutes and that all plots show the values of the variation of the orbital parameters normalized by the fractional scalar field mass $\Lambda$.}
\label{fig:big-picture-orb-elem}
\end{figure*}

\subsection{Using a fixed direction of the black hole spin}

From Appendix~\ref{app:dkdt-orbital-parameters}, one can see that the derivatives of the functions $P_1(r)$ and $P_2(r)$ only influence the radial force and that the function $P_1(r)$ does not participate in the calculations; we use those expressions in the integral of Eq.~\eqref{eq:Delta-kappa} and we are able to calculate the average value of variation of the orbital parameters as a function of the mass coupling parameter $\alpha$. We present those results in Fig.~\ref{fig:big-picture-orb-elem} considering that the unperturbed orbit is characterized by the orbital parameters of Eq.~\eqref{eq:orbital-parameters}. The results are scaled to the relative mass of the scalar field, $\Lambda$. One can see that the values of the factor $\alpha$ that give rise to large variations of the orbital parameters are, approximately, in the range
\begin{equation}
  0.001 \lesssim \alpha \lesssim 0.05,
\end{equation}
which correspond to
\begin{equation}
    1.2\times 10^4\lesssim R_\mathrm{peak}\lesssim 3\times 10^6.
\end{equation}
This range of $\alpha$ is comparable with the orbital range of S2 ($3\times 10^3\lesssim r \lesssim 5\times 10^4$). As expected S2 dynamics is mostly altered when it crosses regions of the scalar field that are associated to relatively high density. Moreover, one verifies that:
\begin{enumerate}

\item For very small and large $\alpha$ the effects are negligible. A very small $\alpha$ corresponds to a location of the cloud far out from the orbit of S2. A large value of $\alpha$ corresponds to a location well inside it, acting like a point source of negligible mass with respect to the black hole (we take $\Lambda \ll 1$).

\item The average variation of the semi-major axis $\langle\Delta a\rangle/\Lambda$ is negligible;

\item There's a maximum value of $\langle\Delta i\rangle/\Lambda$, $\langle\Delta e\rangle/\Lambda$ and $\langle\Delta \Omega\rangle/\Lambda$ and a minimum of $\langle\Delta \omega\rangle/\Lambda$. The maximum of the first three elements occur for the same value of $\alpha \sim 0.012$, while the minimum of $\langle \Delta \omega\rangle/\Lambda$ occurs for $\alpha \sim 0.022$.

\item The variations $\langle \Delta i \rangle/\Lambda$ and $\langle \Delta e \rangle/\Lambda$ may present a positive or negative variation depending on the mass of the scalar field. Their dependence on $\alpha$ is the same and for the value of $\alpha \sim 0.022$ it is observed that $\langle \Delta i \rangle = \langle \Delta e \rangle = 0$ (notice that for the same value of $\alpha$, $\langle \Delta \omega\rangle/\Lambda$ attains its minimum value). For mass coupling parameters $\alpha > 0.022$, the variation of these elements is negative.

\item The angular parameters present variations with different orders of magnitude. The smallest is the variation of the inclination, then the longitude of the ascending node, the argument of the pericentre and the largest corresponds to the variation of the mean anomaly at epoch $\mathcal{M}_0$.

\end{enumerate}

In order to compare the scalar field cloud results with other predictions, we have to make an assumption on the value of the parameter $\Lambda$; we will make the conservative assumption of $\Lambda = 0.01$, i.e., the mass of the scalar field cloud is 1\% of the mass of the central black hole in agreement with the current $\sim 1\%$ upper limits\\ \citep{Gillessen2009,Gravity2018a}. Having established this, we will turn to the plots to obtain the following orders of magnitude for the change in the orbital parameters per orbit
\begin{equation}
  \begin{cases}
    \langle\Delta a\rangle \sim 10^{-9}\\
    \langle\Delta e\rangle \sim \pm 10^{-4}\\
    \langle\Delta i\rangle \sim \pm 0.01'\\
    \langle\Delta \Omega\rangle \sim 0.1'\\
    \langle\Delta \omega\rangle \sim - 10'\\
    \langle\Delta \mathcal{M}_0\rangle \sim 200' 
  \end{cases}
\end{equation}
which support the case that the effects due to the scalar field cloud are comparable to the effects due to the static component of the first post-Newtonian correction.


\subsection{Varying the orientation of the black hole spin}

\begin{figure*}
\includegraphics[width = 2\columnwidth]{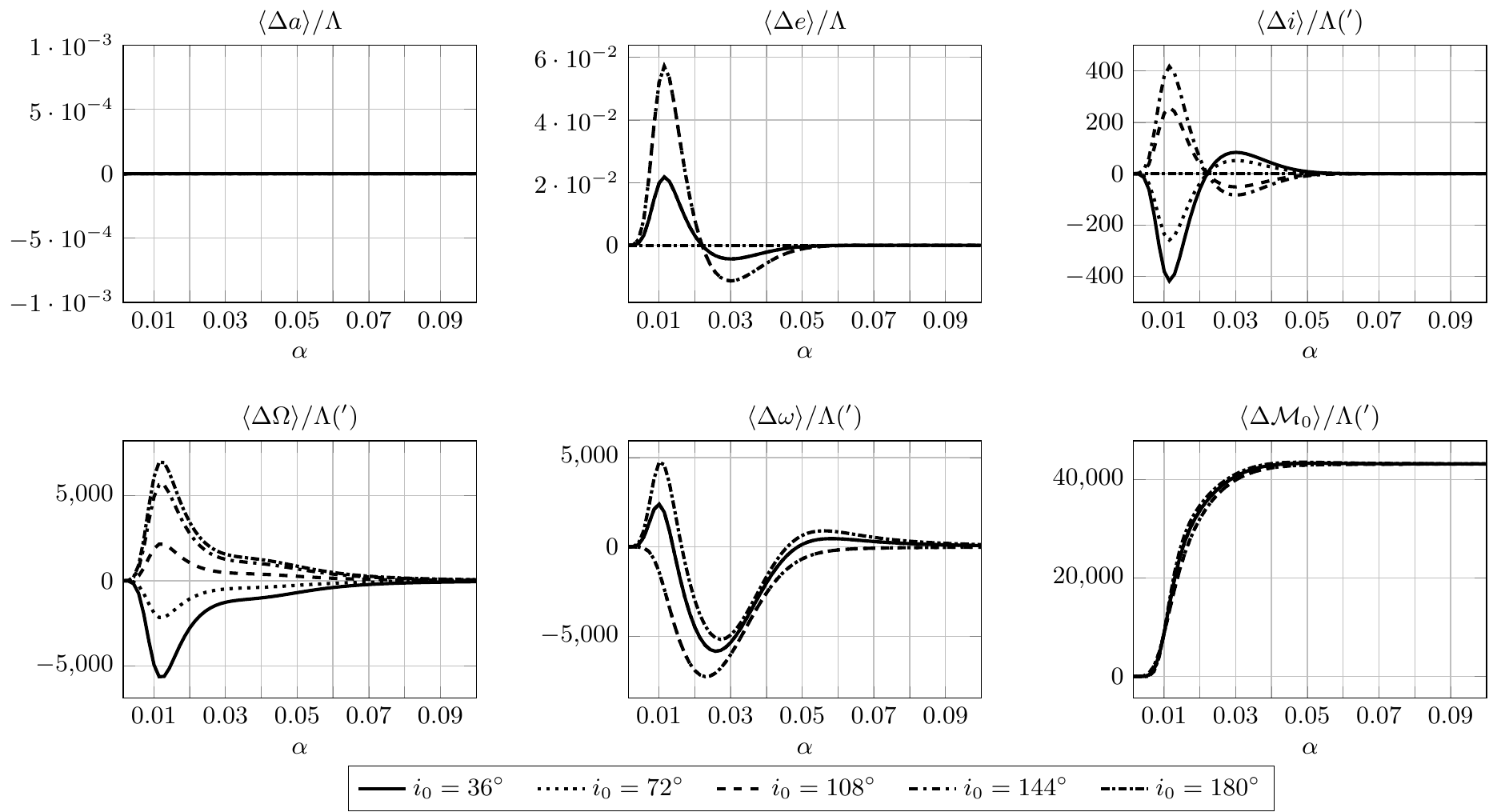}\\
\caption{Redoing the calculations presented in Fig.~\ref{fig:big-picture-orb-elem} using the orbital elements in Eq.~\eqref{eq:orbital-parameters} except for the value of inclination $i_0$, which is varied. Note that for some plots the curves are superimposed.}
\label{fig:big-picture-orb-elem-varying-INC}
\end{figure*}

\begin{figure*}
\includegraphics[width = 2\columnwidth]{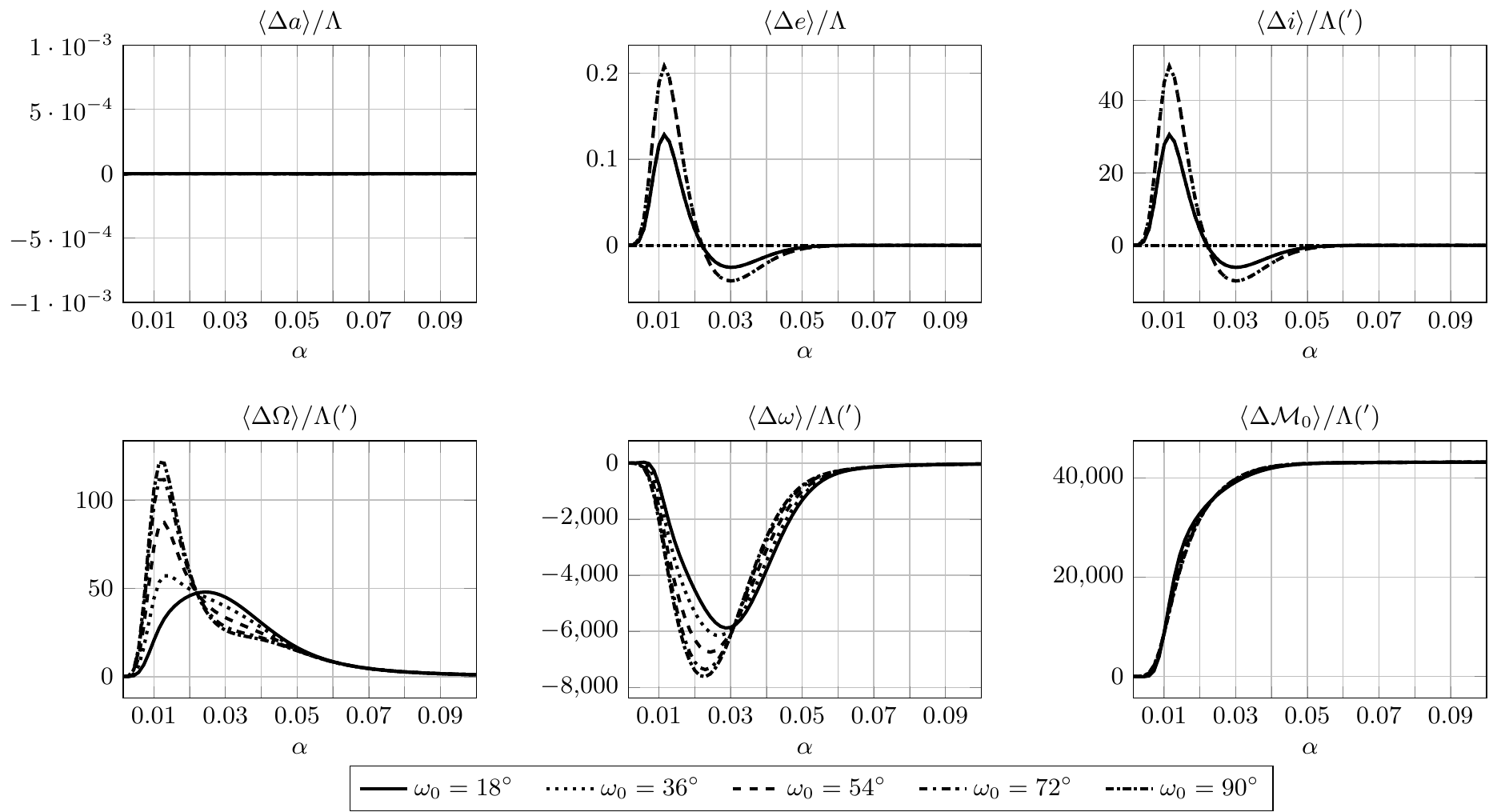}\\
\caption{Redoing the calculations presented in Fig.~\ref{fig:big-picture-orb-elem} using the orbital elements in Eq.~\eqref{eq:orbital-parameters} except for the value of the argument of the pericenter $\omega_0$, which is varied.}
\label{fig:big-picture-orb-elem-varying-SMALLo}
\end{figure*}

Given the uncertainty in the orientation of the black hole spin we can argue that the the orbital parameters with respect to the black hole-centered frame of reference, Eq.~\eqref{eq:orbital-parameters}, cannot be considered with certainty either. This means that one should explore the range of values that one can assign to them. We point out that the calculation of $\langle\Delta\kappa\rangle$ does not depend on the orbital parameter $\Omega_0$, so, we will focus only on $i_0$ and $\omega_0$. The results are presented in Fig.~\ref{fig:big-picture-orb-elem-varying-INC} and Fig.~\ref{fig:big-picture-orb-elem-varying-SMALLo}.
%
%

The conclusions one can take from observing the plots with the varying values of the inclination angle $i_0$ and the argument of the pericentre $\omega_0$, is that the orbital changes  are much more sensitive to the former than to the latter. There are, however, two points in common between the two cases: the variation of the semi-major axis remains negligible, such that one can say that $\langle a \rangle \approx 0$, and the variation of the mean anomaly at epoch is, to all purposes, unaffected by the different values of $i_0$ and $\omega_0$.

In Fig.~\ref{fig:big-picture-orb-elem-varying-INC}, for different values of the initial inclination $i_0 \in ]0,\pi[$, we observe a significant change in the profile of the relations $\Delta\kappa \text{ vs. }\alpha$:
\begin{enumerate}
\item The variation of the eccentricity remains, similarly to the case of Fig.~\ref{fig:big-picture-orb-elem}, negligible.
\item The variation of inclination and longitude of the ascending node are significantly affected by the inclination of the orbit. One can see that the profile of dependence of these two quantities on $\alpha$ changes both in order of magnitude and in sign. For instance, $\langle \Delta \omega\rangle/\Lambda$ is, independently of the value of $\alpha$, always positive if $i_0 = 144^{\circ}$ and always negative if $i_0 = 36^{\circ}$;
\item We verify that for some values of the parameters $\alpha$ and $i_0$, the variation of $\omega$ is positive, which is not verified in Fig.~\ref{fig:big-picture-orb-elem}. Besides this new feature, the order of magnitude of the effect does not change with respect to reference case of Fig.~\ref{fig:big-picture-orb-elem}.
\end{enumerate}
A consequence of the uncertainty in the orbital parameter $i_0$ is the widening of the range of possible values for the variation of the orbital parameters due to the presence of the scalar cloud. Assuming, again, that $\Lambda = 0.01$, the orders of magnitude for the variation of each of the orbital parameters can reach up to
\begin{equation}
  \begin{cases}
    \langle\Delta a\rangle \sim 10^{-9}\\
    \langle\Delta e\rangle \sim \pm 10^{-4}\\
    \langle\Delta i\rangle \sim \pm 1'\\
    \langle\Delta \Omega\rangle  \sim \pm 10'\\
    \langle\Delta \omega\rangle \sim \pm 10'\\
    \langle\Delta \mathcal{M}_0\rangle \sim 100'  
  \end{cases}
\end{equation}
depending on the value of the initial inclination $i_0$.

From  Fig.~\ref{fig:big-picture-orb-elem-varying-SMALLo}, where we present the results of varying the value of $\omega_0$, we observe a much weaker influence of such variation in the shape and order of magnitude of the profiles $\Delta\kappa \text{ vs. }\alpha$: 
\begin{enumerate}

\item One verifies that some values of $\omega_0$, the variation of the eccentricity can be one order of magnitude bigger than that of Fig.~\ref{fig:big-picture-orb-elem}. However, and given that $\Lambda$ is expected to be very small, one can conclude that no matter the actual value of $\omega_0$, the contribution of the scalar cloud to the variation of the eccentricity will always be negligible;

\item The influence of the value of $\omega_0$ to the variation $\langle \Delta i \rangle$ is significant because, although one doesn't verify that the maximum possible value of $\langle \Delta i \rangle$ changes, one sees that, for certain  value of $\omega_0$, the variation of the inclination reduces to zero.

\item With respect to the variations $\langle \Delta \Omega \rangle$ and $\langle \Delta \omega \rangle$, one observes that different values of $\omega_0$ have no significant influence on them, except that it may suppress the magnitude of these variations with respect to Fig.~\ref{fig:big-picture-orb-elem}.

\end{enumerate}
Different values of $\omega_0$ do not introduce much change in the orders of magnitude of the potential effects of the scalar field cloud on the orbital parameters of the orbit. In fact, an inspection of Fig.~\ref{fig:big-picture-orb-elem-varying-SMALLo}, considering $\Lambda = 0.01$, is translated in
\begin{equation}
  \begin{cases}
    \langle\Delta a\rangle \sim 10^{-9}\\
    \langle\Delta e\rangle \sim \pm 10^{-3}\\
    \langle\Delta i\rangle \sim \pm 0.1'\\
    \langle\Delta \Omega\rangle \sim 0.1'\\
    \langle\Delta \omega\rangle \sim - 10'\\
    \langle\Delta \mathcal{M}_0\rangle \sim 100',
  \end{cases}
\end{equation}
which is very similar to the reference case of Fig.~\ref{fig:big-picture-orb-elem}.

\section{Discussion}\label{sec:discussion}

The results obtained in Section~\ref{sec:results} should be compared with the largest relativistic effect on the orbit of S2, which is due to the static component of the first Post-Newtonian correction, and produces the advance of the pericentre, given by \citep[e.g.,][]{will2008testing,preto2009post,Iorio2017a}
\begin{equation}
  \langle \Delta \omega \rangle = \frac{6\pi}{a(1-e^2)} \sim 11'.
\end{equation}
The contribution of the scalar field cloud to $\langle \Delta \omega \rangle$, in the  conservative assumption made in Section \ref{sec:results} is of the order of the GR effect and may be large enough to be detected by GRAVITY. Its contribution can reinforce or reduce the GR value, depending on the black hole spin. By combining the current upper limits of \cite{Hees2017, Parsa2017} for the pericentre shift with the model predictions the fractional mass of the scalar cloud is constrained to  $\lesssim 1 \%$, for the $\alpha$ range with strongest effects. New measurements by Gravity et al. (2019, in preparation) are expected to put stronger constrains.

Other contributions are expected to the pericentre shift, as stressed by \cite{will2008testing}. Second Post-Newtonian order effects, tidal distortions of the star near the pericentre or an extended distribution of mass inside  its orbit are expected to influence the amount of variation of the pericentre longitude. Following the treatment by \cite{preto2009post, merritt2010testing, amaro2019x}, one can calculate the average variation of the orbital parameters of the S2 star as a result of the presence of an extended, power-law, mass distribution of stars (characterized by a exponent $\gamma$) that generates an average potential (see Appendix \ref{app:galactic-potential}). Considering two extreme cases -- a "light" and a "heavy" case corresponding to two limits of the total mass of the extended mass distribution -- one finds, for the light case $\gamma = 1.5$, $\langle\Delta\omega\rangle \sim - 1.37'$ and for the heavy case $\gamma = 2.1$, $\langle\Delta\omega\rangle \sim - 17.19'$. These results indicate that this effect can be competitive with the first post-Newtonian correction with respect to the argument of the pericentre. 

A distinctive aspect of the scalar field is its lack of spherical symmetry that translates in non-null $\langle\Delta i\rangle$ and $\langle\Delta \Omega\rangle$ whose intensity depends on the black hole inclination. The GR-predicted frame-dragging effects \citep[e.g.,][]{will2008testing} will depend on the magnitude and direction of the spin of the black hole. One can have an estimate of these values by considering that the direction of the black hole's spin maximizes the respective contributions, which are constrained from above by (see \cite{Iorio2017b})
\begin{align}
    \langle \Delta i \rangle &\lesssim \frac{4\pi \chi}{n a^3 (1-e^2)^{3/2}} \sim 0.1'\chi \\
    \langle \Delta \Omega \rangle &\lesssim \frac{4\pi \chi}{n a^3 (1-e^2)^{3/2}} \frac{1}{\sin i} \sim 0.1'\chi
\end{align}
where $\chi \equiv (c/G)(S_{\bullet}/M_{\bullet}^2)$ is the dimensionless angular momentum parameter of the black hole and must be smaller than 1 because of the Cosmic Censorship Conjecture. For some black hole inclinations the scalar field effects are larger than the GR ones. Furthermore, the presence of the scalar-field cloud also induces variations in the eccentricity that are near the current precision of \cite{Gravity2018a,Gravity2019} -- future measurements by GRAVITY are expect to place further constraints on this parameter. Although naive, these estimates show that the presence of a scalar field cloud in the vicinity of the black hole in the Galactic Centre may be detectable through the deviations of the variations of the orbital parameters with respect to the GR-predicted values.

The other S-stars have semi-major axis values in the same order of magnitude as the S2-star. This means that a scalar cloud that affects the latter will also affect the other S-stars. Adding to this the fact that the other S-stars have different angular orbital parameters and the fact that, according to Fig.~\ref{fig:big-picture-orb-elem-varying-INC}, the value of the inclination of the orbit can produce a big change in the order of magnitude of the variation of the orbital parameters, a careful study of all the S-stars will be a robust test on the hypothesis of the scalar field cloud.

Finally, the $\alpha$ range can be translated to a scalar field mass parameter in the range
\begin{equation}
  10^{-20}\text{ eV}/c^2 \lesssim m_s \lesssim 10^{-18}\text{ eV}/c^2.
\end{equation}
For comparison, the upper bound on the photon's mass is $10^{-18}\text{ eV}/c^2$ \citep{tanabashi2018review}. \cite{hui2017ultralight} find that cosmological dark matter with energies $10^{-21}\text{ eV}/c^2$ are favoured by observations. This range could be probed by stars further out than S2, provided accurate Newtonian corrections are measurable. Furthemore, the discovery of putative closer stars \citep{Waisberg2018} would constrain the scalar field distributions nearer  the black hole.

\section{Conclusions}\label{sec:conclusion}

The possibility of testing the presence of scalar field structures around black holes has received a lot of attention, particularly with the future gravitational-wave detector facilities. In fact, the dynamics of Extreme-Mass-Ratio-Inspirals -- which will be one of target-systems of LISA -- provide a good way to test the existence of such structures \citep[e.g.,][]{Hannuksela2019, ferreira2017orbital}. Following this trend, we considered the hypothesis that such a scalar field structure may be associated with the black hole in the centre of our galaxy. Using standard results of orbital perturbation theory, we computed how much the orbital parameters of the S2 change over an orbital period. The S2 star will only be sensitive to a scalar field whose mass sits in the range $10^{-20}\text{ eV} \lesssim m_s \lesssim 10^{-18}\text{ eV}$, but if that is the case, then the scalar field cloud will have an effect on orbits orientation which, even in the conservative case of a scalar cloud with 1\% of the black hole mass, can have a significant effect, detectable by GRAVITY. A detailed fit of the available and forthcoming pericentre shift data is addressed in future work.

The plausibility of these scalar field clouds typically depends on the spontaneity of their growth by superradiance mechanism. This is certainly one of the most natural and Occam's razor-friendly ways of realizing the existence of such an astrophysical structure. Furthermore, numerical studies have shown \citep[see e.g.,][]{dolan2013superradiant,witek2013superradiant,okawa2014black} that black holes may support scalar field structures that resulted from a scattering event. In this scenario, a scalar wave meets a black hole and part of it gets trapped in the quasi-bound state structure of the hole. Putting this possibility along with the superradiant mechanism -- which would ``feed'' the trapped scalar field structure --  and we have stronger reasons to consider that black holes harbouring scalar field clouds is an hypothesis worth considering.

\section*{Acknowledgements}

We are very grateful to our funding agencies (MPG, ERC, CNRS, DFG, BMBF, Paris Observatory, Observatoire des Sciences de l'Univers de Grenoble, and the Funda{\c{c}}{\~a}o para a Ci{\^e}ncia e Tecnologia), to ESO and the ESO/Paranal staff, and to the many scientific and technical staff members in our institutions who helped to make GRAVITY a reality.
We would like to thank V\'itor Cardoso and Carlos A. R. Herdeiro for fruitful discussions.
This research was partially supported by Funda\c{c}\~ao para a Ci\^{e}ncia e a Tecnologia, with grants reference UID/FIS/00099/2013, SFRH/BSAB/142940/2018 (P.G.). M.F. acknowledges financial support provided by Funda\c{c}\~ao para a Ci\^{e}ncia e Tecnologia Grant No. PD/BD/113481/2015 awarded in the framework of the Doctoral Programme IDPASC-Portugal.  S.G. acknowledges support from ERC starting grant No. 306311 (PROGRESO). F.E. and O.P. acknowledge support from ERC synergy grant No. 610058 (BlackHoleCam). J.D., M.B., and A.J.-R. were supported by a Sofja Kovalevskaja award from the Alexander von Humboldt foundation




\bibliographystyle{mnras}
\bibliography{orbitsBIB} 




\appendix

\section{Black Hole - Scalar Field setup}
\label{app:BH-SF-setup}

Considering a spacetime described by a fixed Kerr metric, the Klein-Gordon equation
\begin{equation}
    \nabla_{\mu}\nabla^{\mu}\Psi = \mu^2\Psi,
\end{equation}
has solutions that inherit the symmetries of the background, more specifically, these solutions have, in Boyer-Lindquist coordinates $(t,r,\phi,\theta)$, the form\footnote{This is the approach we are going to follow, which is explained in more detail in \cite{detweiler1980klein,rosa2010extremal,yoshino2014gravitational,brito2015black}.} 
\begin{equation}\label{eq:Psi_modes}
  \Psi = \mathrm{e}^{-\mathrm{i} \omega t + \mathrm{i} m \phi}S_{\ell m}(\theta) \psi_{\ell m}(r).
\end{equation}
where $\ell,m$ correspond to angular modes and $\omega$ is the frequency of the field. This expression is substituted in the Klein-Gordon equation and with the resulting expression the bound state spectrum is obtained\footnote{This corresponds to imposing bound state boundary conditions to the Klein-Gordon equation, i.e. looking for solutions such that close to the horizon the solution is an ``ingoing wave'' and that at infinity it describes an exponential decay.}. The allowed values of $\omega$, which are complex numbers, will depend on the parameters of the system and, typically, they are found by solving the Klein-Gordon equation numerically \citep[see][and references therein]{brito2015black}. However, for small mass coupling $\alpha$ (i.e. for the case in which the Compton wavelength of the scalar field is much larger than the gravitational radius of the black hole), it is shown in  \cite{detweiler1980klein} that the bound state frequencies are given by $\omega = \omega_R + \mathrm{i}\omega_I$ with 
\begin{equation}
\label{eq:frequencies-planck-units}
  \begin{cases}
    \omega_R &\sim m_s - m_s \left(\frac{\alpha}{\ell + n +1}\right)^2,\\
    \omega_I &\sim m_s\left(\frac{a_{BH} m}{M_{\bullet}} - 2m_s r_+\right) \frac{\alpha^{4\ell +4}}{\sigma_{\ell}},
  \end{cases}
\end{equation}
in which $r_+ = M_{\bullet} + \sqrt{M_{\bullet}^2 - a_{BH}^2}$, with $a_{BH}$ being related to the black hole angular momentum $J = a_{BH} M$ and $\sigma_{\ell}$ representing a value that depends on the parameters of the system \citep[see, e.g.,][]{brito2015black} and
\begin{equation}
  \begin{cases}
    \psi_{\ell n}(r) &= A_{\ell n} v^{\ell}\mathrm{e}^{-v/2} L_n^{2\ell +1}(v),\\
    S_{\ell m}(\theta) &= P^{m}_{\ell}(\cos\theta),
  \end{cases}
\end{equation}
where $n$ is an integer that identifies the solution\footnote{Similar to the radial quantum number in the analytical solution of the orbitals of a hydrogen atom \citep[e.g.,][]{greiner2011quantum}} , $A_{\ell n}$ is a normalization constant, $P^{m}_{\ell}$ are associated Legendre polynomials, $L_n^{2\ell + 1}$ are generalized Laguerre polynomials and
\begin{equation}
  v = \frac{2 r M_{\bullet} m_s^2}{\ell + n +1},
\end{equation}
with $r$ being the Boyer-Lindquist radial coordinate in Planck units. Notice that the fact that the values of the frequency have a imaginary part implies that with time, the value of the scalar field will grow if $\omega_I > 0$. The growth of this scalar cloud is a consequence of superradiance \citep{brito2015superradiance}. The modes of the bound states solutions which satisfy the superradiance condition, $\omega_R < m\Omega$ where $\Omega = a_{BH}/(r_+^2 + a_{BH}^2)$, will extract energy from the black hole which, since they are bounded, remain localized (instead of being radiated to infinity) in the vicinity of the black hole. It is this process that justifies the presence of an imaginary part of the frequency, giving rise to an exponential growth of the amplitude of the bound states.

The field mode that will grow more efficiently due to the superradiant mechanism is the $n=0, \ell = m = 1$ mode \cite{brito2015black} for which the scalar field function can then be written as
\begin{equation}\label{eq:Psi}
  \Psi = \left[A_{10}\mathrm{e}^{\omega_I t}\right] \mathrm{e}^{-\mathrm{i} (\omega_R t - \phi)}  r (M_{\bullet}m_s^2) \mathrm{e}^{-\frac{r (M_{\bullet}m_s^2)}{2}} \sin\theta.
\end{equation}
One can normalize the coordinates in terms of the black hole mass by applying the substitution
\begin{equation}
  \label{eq:normalized-coordinates}
  \begin{cases}
    r &\rightarrow \bar{r} = r/M_{\bullet}\\
    t &\rightarrow \bar{t} = t/M_{\bullet}
  \end{cases},
\end{equation}
such that the scalar field function is written as
\begin{equation}
  \Psi =  A_0\mathrm{e}^{-\mathrm{i} (\bar{\omega}_R \bar{t} - \phi)}  \bar{r} \alpha^2 \mathrm{e}^{-\frac{\bar{r} \alpha^2}{2}} \sin\theta,
\end{equation}
with $A_0$ denoting the term in square brackets in equation~\eqref{eq:Psi} (which due to the difference between $\omega_R$ and $\omega_I$ can be considered as a constant) and making the dependence on the mass coupling parameter, $\alpha$, explicit. Notice also that with the normalized coordinates, the frequencies are measured in units of $M_{\bullet}^{-1}$ -- we identify that fact by a bar over the corresponding symbol -- meaning that we can write
\begin{equation}
    \begin{cases}
    \bar{\omega}_R &\sim \alpha - \alpha \left(\frac{\alpha}{\ell + n +1}\right)^2\\
    \bar{\omega}_I &\sim \left(\frac{a}{M_{\bullet}}m - 2\alpha \bar{r}_+\right) \frac{\alpha^{4\ell +5}}{\sigma_{\ell}}
  \end{cases}.
\end{equation}
Thus obtaining equation~\eqref{eq:Psi_main}.

\section{The Poisson equation} \label{app:einstein-to-poisson}

In this appendix we reintroduce the fundamental constants. We start with the Einstein equations:
\begin{equation}
  G_{\alpha\beta}  = \frac{8 \pi G}{c^4} T_{\alpha\beta},
\end{equation}
where the Einstein tensor $G_{\alpha\beta}$ reads
\begin{equation}
  G_{\alpha\beta} = R_{\alpha \beta} - \frac{1}{2} R g_{\alpha \beta}.
\end{equation}
In a portion of spacetime that is approximately flat, one can assume that the metric tensor is written as \citep{poisson2014gravity}
\begin{equation}
  \begin{cases}
    g_{00} &= -1 + \frac{2}{c^2} U + \mathcal{O}(c^{-4})\\
    g_{0j} &= \mathcal{O}(c^{-3})\\
    g_{jk} &= \left(1  + \frac{2}{c^2}U \right) \delta_{jk} + \mathcal{O}(c^{-4})
  \end{cases}.
\end{equation}
In this case, the  Einstein tensor reads
\begin{equation}
  \begin{cases}
    G_{00} &=-\frac{2}{c^2}\nabla^2U + \mathcal{O}(c^{-4})\\
    G_{0j} &= \mathcal{O}(c^{-3})\\
    G_{jk} &=  \mathcal{O}(c^{-4})
  \end{cases}.
\end{equation}

The energy momentum tensor we're considering is the one generated by a scalar field, i.e.,
\begin{equation}
  T_{\mu\nu} = \frac{1}{2} \Big[ \Psi_{,\mu} \Psi^*_{,\nu} + \Psi_{,\nu} \Psi^*_{,\mu} - g_{\mu\nu} \Big(\Psi^{,\sigma}\Psi^*_{,\sigma} + \frac{m_S^2 c^2}{\hbar^2} |\Psi|^2\Big)\Big],
\end{equation}
whose components can be organized as
\begin{align}
  \frac{1}{c^4} T_{00} &=  \left(\frac{m_S^2}{2 c^2 \hbar^2}|\Psi|^2  + \frac{1}{2}(\partial_0\Psi)(\partial_0\Psi^*)\frac{1}{c^4} + \mathcal{O}(c^{-4})\right),\\
  \frac{1}{c^4} T_{0j} &= \mathcal{O}(c^{-5}),\\
  \frac{1}{c^4} T_{jk} &= \left(-\frac{m_S^2}{2 c^2 \hbar^2}|\Psi|^2 + \frac{1}{2}(\partial_0\Psi)(\partial_0\Psi^*)\frac{1}{c^4}  + \mathcal{O}(c^{-4})\right),
\end{align}
where $\partial_0 \equiv \partial/\partial(ct)$ in which $t$ is the coordinate time.

The scalar field we are considering is given by (we are using the fastest growing mode $\ell = m = 1, n=0$)
\begin{equation}
  \Psi = A_0 \mathrm{e}^{-\mathrm{i}(\omega_R t - \phi)} r (r_g \mu^2)\mathrm{e}^{-\frac{r  (r_g \mu^2)}{2}} \sin \theta,
\end{equation}
where we know (see Eq.~\eqref{eq:frequencies-planck-units})
\begin{equation}
  \omega_R = \frac{c^2}{\hbar}\left[m_s - \frac{m_s}{2}\left(\frac{\alpha^2}{2}\right)\right],
\end{equation}
where we added the fundamental constants. Taking this into account, we can schematically rewrite the scalar field function as
\begin{equation}
  \Psi(t,\vec{x}) = \exp\left(-\mathrm{i} \frac{m_sc^2}{\hbar} t \right) \psi(t,\vec{x}),
\end{equation}
where $\psi(t,\vec{x})$ constains not only the spatial dependence of the field but also the dependence on time related to subdominant component of the frequency, i.e.
\begin{equation}
  \psi(t,\vec{x}) \sim \exp\left(\mathrm{i} \frac{m_sc^2}{2\hbar}\left(\frac{\alpha^2}{2}\right) t \right).
\end{equation}
Using this schematic form of the scalar field, one can write the derivatie terms in the energy-momentum tensor as (remember that $\partial_0 \equiv \partial/\partial(ct)$
\begin{equation}
  \frac{1}{2}(\partial_0\Psi)(\partial_0\Psi^*) = \frac{1}{2 c^2} \left(\frac{m_s^2 c^4}{\hbar^2} \Psi \Psi^* + \partial_t\psi \partial_t\psi\right).
\end{equation}
Plugging this result in the expression for the energy-momentum tensor, we obtain  
\begin{align}
  \frac{1}{c^4} T_{00} &= 1 \left(\frac{m_s^2}{2 c^2 \hbar^2}|\Psi|^2  + \frac{1}{2 c^2} \left(\frac{m_s^2 c^4}{\hbar^2} \Psi \Psi^* + \partial_t\psi \partial_t\psi\right)\frac{1}{c^4} + \mathcal{O}(c^{-4})\right),\\
  \frac{1}{c^4} T_{0j} &= \mathcal{O}(c^{-5}),\\
  \frac{1}{c^4} T_{jk} &= \left(-\frac{m_s^2}{2 c^2 \hbar^2}|\Psi|^2 + \frac{1}{2 c^2} \left(\frac{m_s^2 c^4}{\hbar^2} \Psi \Psi^* + \partial_t\psi \partial_t\psi\right)\frac{1}{c^4}  + \mathcal{O}(c^{-4})\right),
\end{align}
which can be simplified to
\begin{align}
  \frac{1}{c^4} T_{00} &= \left( \left[\frac{m_s^2}{2 c^2 \hbar^2} + \frac{m_s^2}{2 c^2 \hbar^2} \right]|\Psi|^2  + \frac{\partial_t\psi \partial_t\psi}{2 c^6} + \mathcal{O}(c^{-4})\right),\\
  \frac{1}{c^4} T_{0j} &= \mathcal{O}(c^{-5}),\\
  \frac{1}{c^4} T_{jk} &= \left( \left[\frac{m_s^2}{2 c^2 \hbar^2} - \frac{m_s^2}{2 c^2 \hbar^2} \right]|\Psi|^2  + \frac{\partial_t\psi \partial_t\psi}{2 c^6} + \mathcal{O}(c^{-4})\right).
\end{align}
Notice that $\partial_t\psi \sim (m_sc^2)/(4\hbar) \alpha^2$ so that
\begin{equation}
  \frac{\partial_t\psi \partial_t\psi}{2 c^6} \sim \frac{m_s^2\alpha^4}{32 \hbar^2c^2}.
\end{equation}
We see, then, that in the limit of validity of the scalar field function we are using -- $\alpha \ll 1$ -- this term is negligible compared to the other terms in $c^{-2}$. So, the energy momentum tensor of the scalar field in the low energy limit is given by
\begin{align}
  \frac{1}{c^4} T_{00} &= \left(\frac{m_s^2}{c^2 \hbar^2} |\Psi|^2 +  \mathcal{O}(c^{-4})\right),\\
  \frac{1}{c^4} T_{0j} &= \mathcal{O}(c^{-5}),\\
  \frac{1}{c^4} T_{jk} &=  \mathcal{O}(c^{-4}).
\end{align}
Finally, we can equate both sides of the Einstein equations
\begin{equation}
  G_{\alpha\beta}  = \frac{8 \pi G}{c^4} T_{\alpha\beta},
\end{equation}
whose 00-component is Poisson equation 
\begin{equation}
  \nabla^2 U = - 4\pi G \rho,
\end{equation}
with
\begin{equation}
    \rho=\left( \frac{m_s^2 c^2}{\hbar^2} |\Psi|^2 \right),
\end{equation}
being the dominant term of $T_{00}$.
\section{Harmonic decomposition} \label{app:solving-the-potential}

\begin{figure*}
  \begin{center}
  \includegraphics[width=1\columnwidth]{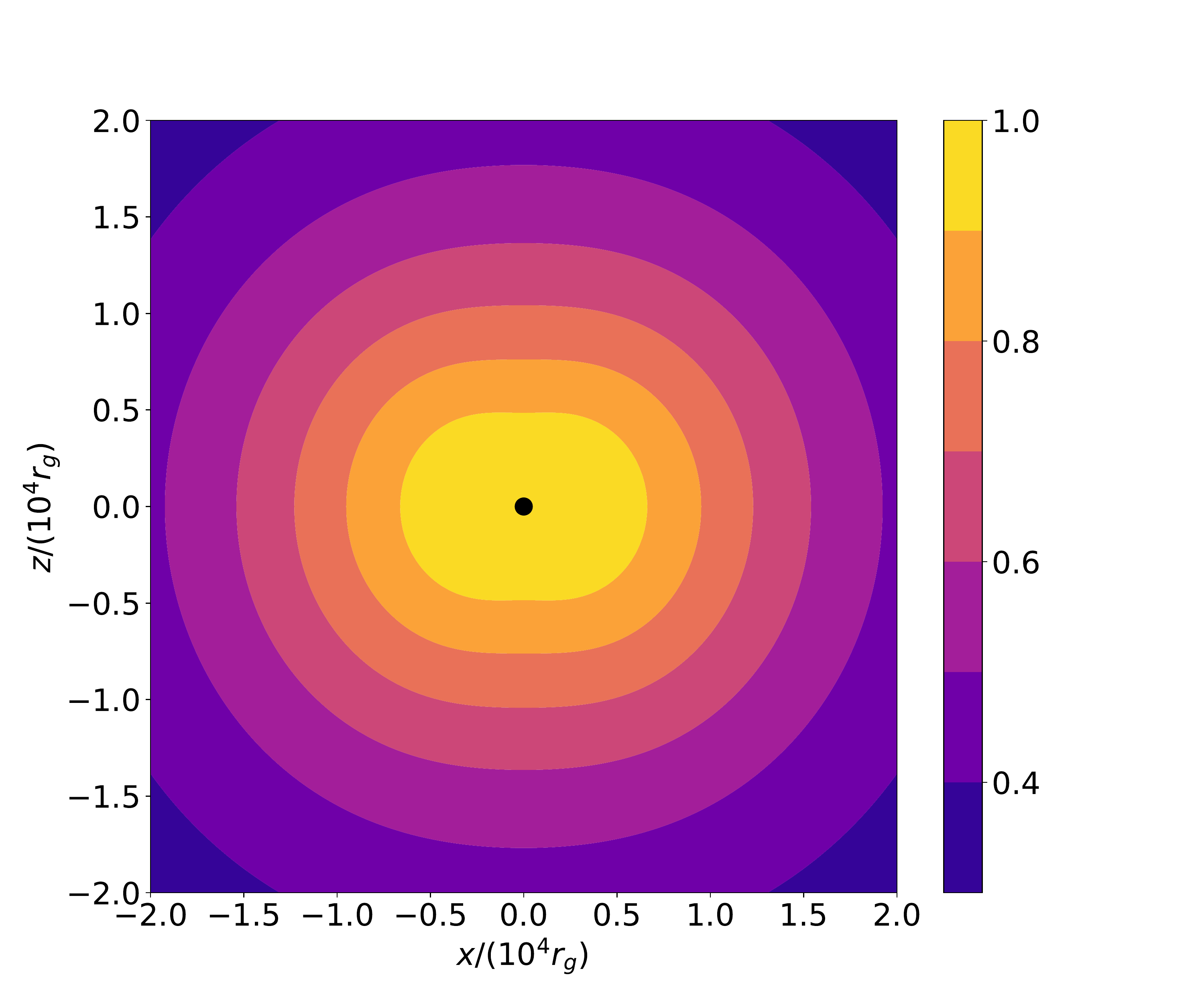}
  \includegraphics[width=1\columnwidth]{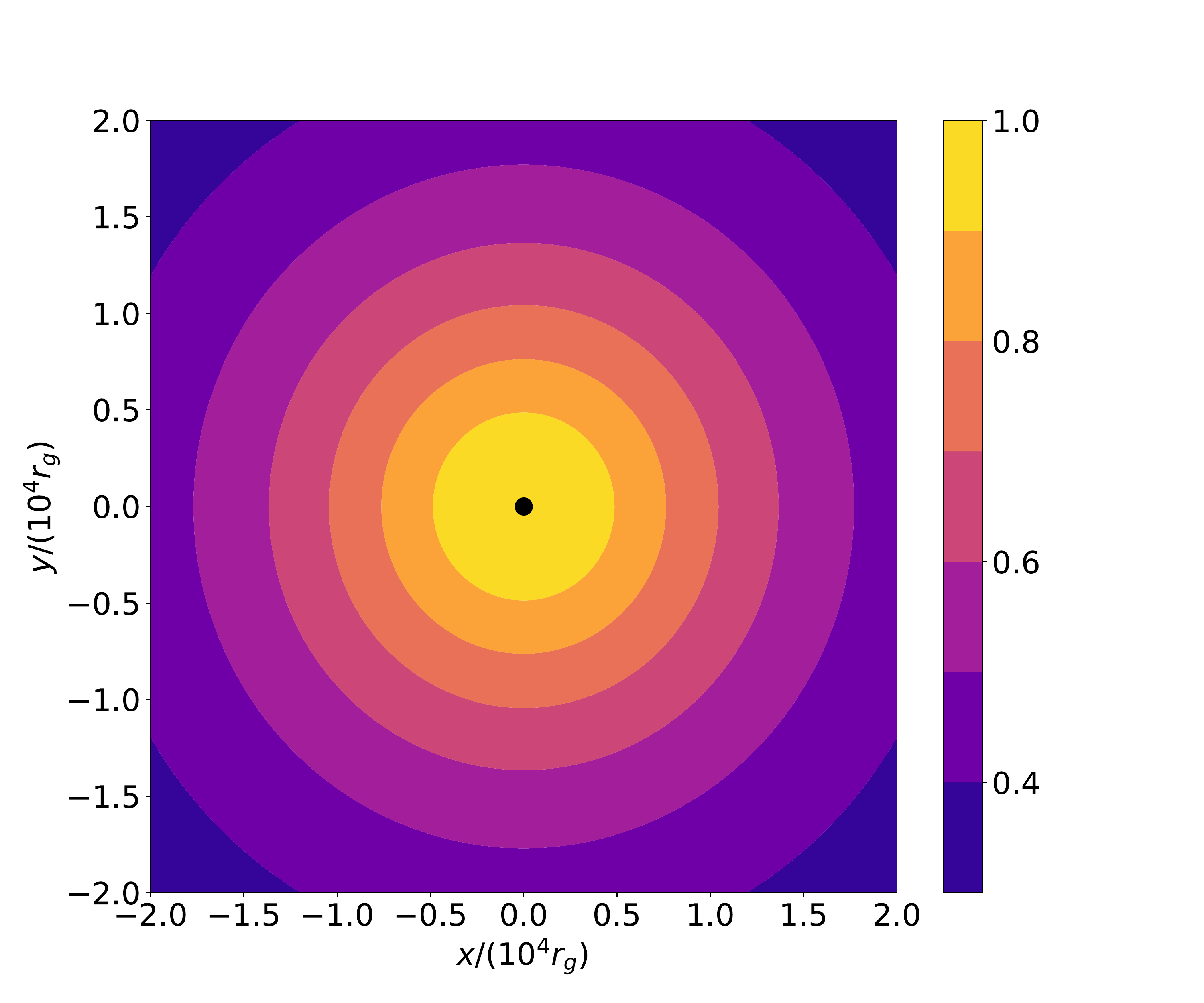}
  \end{center}
  \caption{Normalized potential ($U_\mathrm{sca}$) distribution of the scalar field cloud,  in the black hole reference frame, for $\alpha=0.02$.  The black dot depicts the black hole position. Left: $xz$ cut across $U_\mathrm{sca}$.  Right: $xy$ cut.}
  \label{fig:scheme_U_sca}
\end{figure*}

\begin{figure}
  \begin{center}
  \includegraphics[width=\columnwidth]{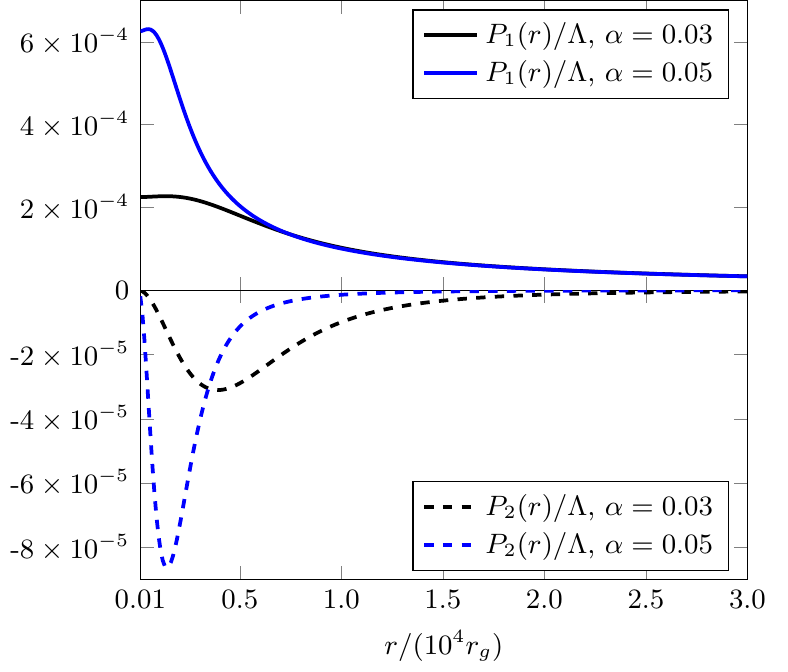}
  \end{center}
  \caption{Potential functions $P_1$ and $P_2$ (see Eqs.~\eqref{eq:potential-function-P1} and \eqref{eq:potential-function-P2}) for two different values of the mass coupling parameter $\alpha$.}
  \label{fig:potential-functions-plot}
\end{figure}

To solve Poisson's equation 
\begin{equation}
  \bar{\nabla}^2 U_{\mathrm{sca}} = - 4\pi \rho
\end{equation}
with (see Eq.~\eqref{eq:poisson-equation-barred-variables})
\begin{equation}
  \rho = \left[\frac{M_{\mathrm{cloud}}}{M_{\bullet}}\right]\left(\frac{\alpha^{10}}{64\pi} \mathrm{e}^{- \alpha^2 \bar{r}} \bar{r}^2 \sin^2\theta\right)
\end{equation}
we employ the spherical harmonic decomposition technique \cite{poisson2014gravity}. With this technique, the solution to the Poisson's equation is given by
\begin{equation}
  \label{eq:potential-sum-terms}
  U_{\mathrm{sca}} = \sum_{\ell m} \frac{4\pi}{2\ell + 1} \left[ q_{\ell m}(r) \frac{Y_{\ell m}(\theta,\phi)}{r^{\ell + 1}} + p_{\ell m}(r) Y_{\ell m}(\theta,\phi)\right]
\end{equation}
where
\begin{equation}
  q_{\ell m}(r) = \int_0^r s^{\ell} \tilde{\rho}_{\ell m}(t,s) s^2 \mathrm{d}s,
\end{equation}
\begin{equation}
  p_{\ell m}(r) = \int_r^{\infty}  \frac{\tilde{\rho}_{\ell m}(t,s)}{s^{\ell+1}} s^2 \mathrm{d}s,
\end{equation}
\begin{equation}
  \tilde{\rho}_{\ell m} = \int \rho Y_{\ell m}^* \sin\theta \mathrm{d}\theta\mathrm{d}\phi.
\end{equation}
We perform these calculations and we verify that only the $(\ell = 0,m = 0)$ and $(\ell = 2,m = 0)$ terms\footnote{These are spherical harmonic terms, not to be confused with the spin weighted modes of equation~\eqref{eq:Psi_modes}.} in the sum of Eq.~\eqref{eq:potential-sum-terms} are different from zero. This means that the gravitational potential can be schematically written as
\begin{equation}
  U_{\mathrm{sca}} = 4\pi \left[\frac{q_{00}}{r}Y_{00} + p_{00}Y_{00}\right] +  \frac{4\pi}{5} \left[\frac{q_{20}}{r^3}Y_{20} + p_{20}r^2Y_{20}\right]
\end{equation}
which can be simplified as
\begin{equation}
  U_{\mathrm{sca}} = \left[\frac{M_{\mathrm{cloud}}}{M_{\bullet}}\right] \left[ P_1(\bar{r}) + P_2(\bar{r})\cos^2\theta\right]
\end{equation}
with
\begin{align}
  P_1(\bar{r}) &= \frac{16 \alpha^4 \bar{r}^2+48}{16 \alpha^4 \bar{r}^3}-\frac{e^{-\alpha^2 \bar{r}}}{16 \alpha^4 \bar{r}^3} \Big[\alpha^{10} \bar{r}^5\nonumber\\
         &+6 \alpha^8 \bar{r}^4+20 \alpha^6 \bar{r}^3+40 \alpha^4 \bar{r}^2+48 \alpha^2 \bar{r}+48\Big]\,,\label{eq:potential-function-P1}\\
  P_2(\bar{r}) &= -\frac{9}{\alpha^4 \bar{r}^3} + \frac{e^{-\alpha^2 \bar{r}}}{16 \alpha^4 \bar{r}^3} \Big[\alpha^{10} \bar{r}^5+6 \alpha^8 \bar{r}^4\nonumber\\
         &+24 \alpha^6 \bar{r}^3+72 \alpha^4 \bar{r}^2+144 \alpha^2 \bar{r}+144\Big]\label{eq:potential-function-P2}.
\end{align}
In Fig.~\ref{fig:scheme_U_sca} $x,y$ and $x,z$ cuts of  $U_{\mathrm{sca}}$ are depicted and in Fig.~\ref{fig:potential-functions-plot}, the functions $P_1$ and $P_2$ are represented for two different values of the mass coupling parameter.

\section{Keplerian orbits formalism} \label{app:keplerian-orbits-formalism}\label{app:dkdt-orbital-parameters}

The starting point of a Keplerian orbit is the equation of motion of a mass in a Keplerian gravitational field; using the re-scaled distance and time coordinates, it reads
\begin{equation}\label{eq:Newton}
  \frac{d^2 \vec{\bar{r}}}{d\bar{t}^2} = -\frac{\vec{\bar{r}}}{\bar{r}^3}.
\end{equation}
where we consider a reference frame centered on the black hole with the z-axis aligned with the angular momentum of the black hole \citep[using the same approach as][]{grould2017general}. In a system of cartesian coordinates, the orbiting mass wil follow a path described by
\begin{equation}
  \label{eq:xyz-orbit}
  \begin{cases}
    x &= r [ \cos\Omega\cos(\omega +f) -\sin\Omega\sin(\omega +f)\cos i]\\
    y &= r [ \sin\Omega\cos(\omega +f) +\cos\Omega\sin(\omega +f)\cos i]\\
    z &= r \sin(\omega + f) \sin i
  \end{cases}
\end{equation}
where
\begin{equation}
  r = \frac{a(1-e^2)}{(1+e\cos f)}.
\end{equation}
The parameters $(a,e,i,\Omega,\omega)$ are the orbital elements characterizing the geometrical shape of the elliptical orbit; the size and the shape are given by the values of the eccentricity ($e$) and the semi-major axis ($a$), the orientation of the orbit is given by the values of the inclination ($i$) and the longitude of the ascending node ($\Omega$); finally the position of the star in the orbit is given by the argument of the pericentre (or periapsis $\omega$) and by the true anomaly ($f$). Only the latter value is not constant for a Keplerian orbit: the true anomaly is $f=0$ when the star is in the pericentre and $f=\pi$ in the apocentre (or apoapsis). The true anomaly is related to the time by the Kepler equation
\begin{equation}
  \mathcal{M} = E - e\sin E
\end{equation}
where
\begin{equation}
  \mathcal{M} = \mathcal{M}_0 + n (t - t_0)
\end{equation}
is the mean anomaly, $\mathcal{M}_0$ is the mean anomaly at epoch\footnote{if we fix $t_0$ to be the time of pericentre passage, then one can set $\mathcal{M}_0 = 0$ implying that $\mathcal{M}(t=t_0 + T/2) = \pi$, which corresponds to the value in the apocentre},
\begin{equation}
  n = \sqrt{\frac{1}{a^3}}
\end{equation}
is the mean motion\footnote{The mean motion is defined by $n=2\pi/T$, where $T$ is the period of the orbit. Before re-scaling, the mean motion reads $n = \sqrt{GM/a^3}$ which becomes $\bar{n} = \sqrt{1/\bar{a}^3}$ after re-scaling.} and $E$ is the eccentric anomaly, which is related to the true anomaly by
\begin{equation}
  \tan \frac{f}{2} = \sqrt{\frac{1+e}{1-e}} \tan \frac{E}{2}.
\end{equation}

The presence of a perturbing force corresponds to, instead of having Eq.~\eqref{eq:Newton}, having Eq.~\eqref{eq:Newton_perturbed} describing the movement of the star. Following the osculating conics method of \cite{kopeikin2011relativistic}, the perturbed orbit will be described by the same expressions of Eq.~\eqref{eq:xyz-orbit} but with orbital parameters varying according to
\begin{align}
  \frac{da}{dt} &= \frac{2}{n\sqrt{1 - e^2}} \left(e F_R \sin f + F_T \frac{p}{r}\right)\label{eq:kappa-1}\\
  \frac{de}{dt} &= \frac{\sqrt{1 - e^2}}{n a} \left[ F_R \sin f + F_T \left( \cos f + \cos E\right) \right]\\
  \frac{d i}{dt} &= \frac{ r \cos(f + \omega)}{n a^2 \sqrt{1 - e^2}} F_N\\
  \frac{d \Omega}{dt} &= \frac{r \sin(f + \omega)}{n a^2 \sqrt{1 - e^2} \sin i} F_N\\
  \frac{d \omega}{dt} &= - \cos i \frac{d\Omega}{dt} + \frac{\sqrt{1-e^2}}{n a e} \left[ - F_R \cos f + F_T \left(1 + \frac{r}{p}\right)\sin f \right] \\
  \frac{d \mathcal{M}_0}{dt} &= - \sqrt{1 - e^2} \left(\frac{d \omega}{dt} + \cos i \frac{d \Omega}{dt}\right) - \frac{2 r}{n a^2} F_R \label{eq:kappa-6},
\end{align}
%
%
%
where $p = a (1 -e^2)$ is the semi latus rectum and
\begin{equation}
  \begin{cases}
    F_R &= \hat{n}\cdot\vec{F}_{\mathrm{pert}}\\
    F_T &= (\hat{k}\times\hat{n})\cdot\vec{F}_{\mathrm{pert}}\\
    F_N &= \hat{k}\cdot\vec{F}_{\mathrm{pert}},
  \end{cases}
\end{equation}
are the radial, transverse and normal (to the orbit) components of the perturbing force. Notice that $\hat{n} = \vec{r}/r$ is the radial unit vector and $\hat{k}$ is the unit vector orthogonal to the instantaneous orbital plane
\begin{equation}
  \hat{k} = \frac{\vec{r} \times \dot{\vec{r}}}{|\vec{r} \times \dot{\vec{r}}|}.
\end{equation}

\subsection*{Equations for each of the orbital parameters} 

Using the perturbing force expression
\begin{equation}
  \vec{F}_{\mathrm{pert}} = \Lambda \nabla \left[ P_1(r) + P_2(r)\cos^2\theta\right]
\end{equation}
to obtain the components
\begin{align}
  F_R/\Lambda &= \sin ^2(i) \sin ^2(f+\omega ) P_2'(r)+P_1'(r)\\
  F_T/\Lambda &=-\frac{\sin ^2(i) (e \cos (f)+1) \sin (2 (f+\omega )) P_2(r)}{a \left(e^2-1\right)}\\
  F_N/\Lambda &=-\frac{\sin (2 i) (e \cos (f)+1) \sin (f+\omega ) P_2(r)}{a \left(e^2-1\right)}
\end{align}
where the prime $ '$ stands for derivative with respect to the radial coordinate, one can rewrite the time derivatives of the orbital parameters as
\begin{align}
  \frac{da}{dt} \frac{dt}{df} &= 2 a^2 (1 - e^2) \left[T_1\right] + 2 a^3 e (1 - e^2) \sin(f) \left[R_2\right]\\
  \frac{d e}{dt} \frac{dt}{df} &= a^2 \left(e^2-1\right)^2 \Bigg(\left(e \cos ^2(f)+e+2 \cos (f)\right) \left[T_3\right] +\nonumber\\
  &+\sin (f) \left[R_2\right]\Bigg)\\
  \frac{d i}{dt} \frac{dt}{df} &=  a^2 \left(e^2-1\right)^2 \cos (f+\omega )\left[N_3\right]\\
  \frac{d \Omega}{dt} \frac{dt}{df} &=  a^2 \left(e^2-1\right)^2 \csc (i) \sin (f+\omega ) \left[N_3\right]\\
  \frac{d \omega}{dt} \frac{dt}{df} =& - \left\{\frac{a^2(e^2-1)^2(-e \sin (f) \cos (f)-2 \sin (f))}{e}\right\} \left[T_3\right]\nonumber\\
                                    & - \left\{\frac{a^2(e^2-1)^2\cos(f)}{e}\right\} \left[R_2\right]\nonumber\\
                                    & - \left\{a^2(e^2-1)^2 \cot(i)\sin(f+\omega)\right\} \left[N_3\right]\\
  \frac{d \mathcal{M}_0}{dt} \frac{dt}{df} =& \frac{a^2 \left(1-e^2\right)^{5/2}}{e} \Bigg(\left(e \cos ^2(f)-2 e+\cos (f)\right)\left[R_3\right] \nonumber\\
                                            & - \sin (f) (e \cos (f)+2)\left[T_3\right]\Bigg)
\end{align}
where
\begin{align}
  T_1 &=-\frac{\sin ^2(i)\sin (2 (f+\omega ))}{a \left(e^2-1\right)}P_2(f),\\
  T_3 &=-\frac{\sin ^2(i)\sin (2 (f+\omega ))}{a \left(e^2-1\right)}\left[\frac{P_2(f)}{(e \cos (f)+1)^2}\right],\\
  N_3 &=-\frac{\sin (2 i) \sin (f+\omega )}{a \left(e^2-1\right)}\left[\frac{P_2(f)}{(e \cos (f)+1)^2}\right],\\
  R_2 &=\left[\frac{P_1'(r)}{(e \cos (f)+1)^2}\right] + \sin^2(i)\sin^2(f+\omega)\left[\frac{P_2'(r)}{(e \cos (f)+1)^2}\right],\\
  R_3 &= \left[\frac{P_1'(r)}{(e \cos (f)+1)^3}\right] + \sin^2(i)\sin^2(f+\omega)\left[\frac{P_2'(r)}{(e \cos (f)+1)^3}\right].
\end{align}

\section{Newtonian effects from extended mass} \label{app:galactic-potential}

One of our assumptions so far is that the S2 star is located far enough from the central black hole that the gravitational effect of the latter on the former can be significantly perturbed by a scalar field cloud. Although we consider that the scalar field cloud effects are cumulative (i.e. they can be added to the other possible effects that may rule the dynamics in the core of the galaxy) it is fundamental that we compare the magnitude of the different effects. In this section we will focus on the effects that arise from the mean gravitational perturbations coming from the other stars or extended mass in the core of the Milky Way.

In spite of the apparent agreement of theoretical studies regarding the distribution of stars in the centre of galaxies -- there are stellar dynamics studies that approached the problem \citep{peebles1972star, bahcall1976star, lightman1977distribution} and N-body confirmation \citep{baumgardt2004massive, preto2004n, freitag2006stellar} -- in which a density distribution of stars around the black hole may be described by a power-law function, describing a stellar cusp around a black hole, the observational results obtained so far appear to validate this theoretical prediction \citep[see][and references therein]{Habibi2019}. A discussion of these matters is beyond the scope of the paper, but in order to get a feeling of the orders of magnitude associated with the effects that may come from the population of stars in the Galactic Centre, we will use the simple approach of modeling the density of the population of stars by a power law \citep[as in][]{preto2009post, merritt2010testing, amaro2019x}. We will consider that the mean density in the Galactic Centre is given by
\begin{equation}
  \rho(r) = \rho_0 \left(\frac{r}{r_0}\right)^{-\gamma}
\end{equation}
which $\rho_0$ the stellar density at the characteristic radius of normalization $r_0$. The enclosed mass, i.e., the mass of the stars that are described by this density function are given by
\begin{equation}
  M(r) = 4\pi\int_{0}^r \rho(x) x^2 dx = \frac{4\pi \rho_0 r_0^3}{3-\gamma} \left(\frac{r}{r_0}\right)^{3-\gamma},\quad \gamma < 3.
\end{equation}
Considering that $r_0 = 0.01\text{ pc}$ the total mass stellar mass within this radius is given by
\begin{equation}
  M_{*}(r_0) = \frac{4\pi \rho_0 r_0^3}{3-\gamma}
\end{equation}
so that we can write that the average galactic potential is given by
\begin{equation}
  U_{\mathrm{gal}}(r) = - \frac{M_{*}(r_0)}{(2 - \gamma) r_0}\left(\frac{r}{r_0}\right)^{2-\gamma }, \quad \gamma \neq 2.
\end{equation}
So, the resulting force that perturbs the Keplerian orbit is ($F = \nabla U_{\mathrm{gal}}$)
\begin{equation}
  F_R = -\frac{ M_{*}(r_0)}{r_0^2}\left(\frac{a-a e^2}{r_0(1 + e \cos (f)}\right)^{1-\gamma }, \quad F_T = F_N = 0.
\end{equation}

Given that \cite{preto2009post} also analyses the S2 star, we are going to follow their choices for the exponents $\gamma$ and the values of the enclosed mass $M_{*}(r_0)$; so, we will consider two cases
\begin{equation}
  \begin{cases}
    \gamma_{l} = 1.5, & M_{*}(r_0) = 2 \times 10^3 M_{\odot}\\
    \gamma_{h} = 2.1, & M_{*}(r_0) = 2 \times 10^4 M_{\odot}
  \end{cases}.
\end{equation}
where the subscript $l$ and $h$ corresponds to the type of stars that are considered to source the density distribution under analysis. Since stars with different masses get distributed with different density profiles, and given the uncertainty associated with modelling the  Galactic potential, these two cases aim to illustrate two extremal cases.

Using the perturbing force due to the average galactic potential, one can calculate the average variation of the orbital parameters of the S2 star; for the light case $\gamma = 1.5$, we obtain 
\begin{equation}
  \begin{cases}
    \langle\Delta a \rangle = \langle\Delta i \rangle = \langle\Delta e \rangle = \langle\Delta\Omega\rangle = 0\\
    \langle\Delta\omega\rangle \sim - 1.37'\\
    \langle\Delta\mathcal{M}_0\rangle \sim 10.31'
  \end{cases}.
\end{equation}
For the heavy case $\gamma = 2.1$, we obtain 
\begin{equation}
  \begin{cases}
    \langle\Delta a \rangle = \langle\Delta i \rangle = \langle\Delta e \rangle = \langle\Delta\Omega\rangle = 0\\
    \langle\Delta\omega\rangle \sim - 17.19'\\
    \langle\Delta\mathcal{M}_0\rangle \sim 103.13'
  \end{cases}.
\end{equation} 


\bsp	
\label{lastpage}
\end{document}